\documentclass[fleqn,usenatbib]{mnras}
\usepackage{newtxtext,newtxmath}
\usepackage[T1]{fontenc}

\usepackage{graphicx}
\usepackage{amsmath}
\usepackage{lipsum}
\usepackage{array}
\usepackage[table]{xcolor}
\usepackage{subcaption}
\usepackage{caption}
\usepackage{bm}
\usepackage{float}
 \usepackage{natbib} 
\usepackage{enumitem}
\usepackage{ulem}
\usepackage{orcidlink}
\usepackage{booktabs}
\usepackage{tabularx}
\usepackage{ragged2e}
\usepackage{makecell}
\usepackage{multirow}
\usepackage{rotating}
\usepackage{graphicx}

\newcommand{\lya}{Ly$\alpha$}

\newcommand{\buv}{$\beta_{\rm UV}$}
\newcommand{\mab}{$M_{\rm UV}$}

\title[Simulating high-z JWST galaxies]{{\bf N}ISER-{\bf I}UCAA {\bf N}ew Simulations of {\bf J}WST G{\bf A}laxies and Quasars (NINJA): Properties of galaxies at $5\le z\le 10$}
\author[Behera et al.]{
    Ranit Behera$^{1}$\thanks{E-mail: ranit.behera@iucaa.in}\orcidlink{0009-0008-6601-3564},
    Raghunathan Srianand$^{1}$\orcidlink{0000-0002-9062-1921},
    Nishikanta Khandai$^{2,3}$\orcidlink{0000-0003-3081-0189},
    Prakash Gaikwad$^{4}$\orcidlink{0000-0002-2423-7905}
    \\
    $^{1}$Inter-University Centre for Astronomy and Astrophysics, Post Bag 4, Ganeshkhind, Pune - 411007, India\\
    $^{2}$School of Physical Sciences, National Institute of Science Education and Research, Jatni 752050, Odisha, India\\
    $^{3}$Homi Bhabha National Institute, Training School Complex, Anushaktinagar, Mumbai 400094, Maharashtra, India\\
    $^{4}$Department of Astronomy, Astrophysics and Space Engineering, Indian Institute of Technology Indore, Simrol, MP 453552, India
}

\date{Accepted XXX. Received YYY; in original form ZZZ}
\pubyear{\the\year{}}

\newcommand{\NINJA}{\textsc{NINJA}}
\newcommand{\ASTRID}{\textsc{ASTRID}}
\newcommand{\BlueTides}{\textsc{BlueTides}}
\newcommand{\EAGLE}{\textsc{EAGLE}}
\newcommand{\IllustrisTNG}{\textsc{IllustrisTNG}}
\newcommand{\MilleniumTNG}{\textsc{MilleniumTNG}}
\newcommand{\THESAN}{\textsc{THESAN}}
\newcommand{\SPHINX}{\textsc{SPHINX}}
\newcommand{\FIRETWO}{\textsc{FIRE-2}}
\newcommand{\FLARES}{\textsc{FLARES}}
\newcommand{\THESANZOOM}{\textsc{THESAN-ZOOM}}
\newcommand{\UniverseMachine}{\textsc{UniverseMachine}}
\newcommand{\MassiveBlackII}{\textsc{MassiveBlack-II}}
\newcommand{\BAHAMAS}{\textsc{BAHAMAS}}
\newcommand{\FLAMINGO}{\textsc{FLAMINGO}}
\newcommand{\SIMBA}{\textsc{SIMBA}}
\newcommand{\COLIBRE}{\textsc{COLIBRE}}

\newcommand{\software}[1]{\texttt{#1}}

\begin{document}
\label{firstpage}
\pagerange{\pageref{firstpage}--\pageref{lastpage}}
\maketitle

\makeatletter
\def\@journal{}
\def\@volume{}
\def\@volumeno{}
\def\@pagerange{}
\def\@pubyear{}
\def\@oddfoot{\hfil}
\def\@evenfoot{\hfil}
\makeatother

\begin{abstract}
    We present the NINJA suite of cosmological hydrodynamical simulations developed to investigate galaxy formation and evolution at \(z \gtrsim 5\) in the era of JWST. Using our fiducial simulation, we {explore} a range of spectral synthesis prescriptions and dust attenuation models, demonstrating that suitably chosen parameters can reproduce the observed UV luminosity functions (UVLFs) over \(5 \leq z \leq 10\). In all cases, the inferred dust-to-metal ratio evolves with redshift, although its normalization at fixed redshift varies by a factor of \(\sim 7\), depending on the adopted dust--metallicity scaling and attenuation curve. These model variations introduce substantial scatter in predictions for the \(B\)-band luminosity function, the H\(\alpha\) luminosity function, the UV slope--UV magnitude relation, the stellar mass--Balmer ratio relation, and the relation between stellar and nebular colour excesses. Simultaneously reproducing these observables across multiple redshifts will therefore be essential for constraining dust models at high redshift with forthcoming observations. Observations of galaxies spanning a broad range of stellar masses with the \mbox{Atacama Large Millimeter/submillimeter Array} (ALMA) will provide particularly strong and independent constraints on dust properties. Our fiducial models underpredict the UV luminosity function at \(z \geq 10\) relative to current observations, even when adopting a top-heavy IMF and neglecting dust attenuation. We find that galaxy properties are not fully converged at these redshifts in our simulation, indicating that higher-resolution simulations are required to robustly model galaxies at \(z > 10\). We further emphasize that degeneracies between feedback prescriptions used in our simulation and dust properties must be carefully addressed when interpreting high-redshift observations and calibrating galaxy formation models.
\end{abstract}


\section{Introduction}
\label{sec:intro}

The \textit{James Webb Space Telescope} \citep[JWST, see][for the scientific objectives]{Gardner2023} provides high spatial and spectral resolution observations, in rest-frame ultraviolet (UV) to optical wavelengths, of faint high-$z$ galaxies responsible for cosmic reionization. Highly sensitive photometric measurements provide galaxy luminosity functions in multiple bands (especially rest-frame far-UV), rest-frame UV slopes (i.e. \buv\ from $f_\lambda \propto \lambda^{\beta_\text{UV}}$ typically over 1300–2800 \AA), real space and projected angular clustering, stellar masses (${\rm M_\star}$) from SED (spectral energy distribution) fitting, galaxy morphology and size, and their redshift evolution. Spectroscopic observations constrain (i) spectral line luminosity functions, (ii) star formation rates (SFRs), (iii) metal abundances and relative ratios of metals, (iv) the  presence and properties of supermassive black holes, (v) properties of the interstellar (ISM) and circumgalactic (CGM) media of galaxies, and (vi) the nature and ionization state of the intergalactic medium.

Observations with the JWST have significantly improved constraints on the ultraviolet luminosity function (UVLF) of galaxies at \( z \geq 7 \) compared to earlier measurements from the \textit{Hubble Space Telescope} (HST) \citep[e.g.][]{Adams2024, Donnan2024, Harikane2025, Weibel2025, Bouwens2015, Bouwens2021, Bouwens2022, Finkelstein2015, Bowler2015, Bowler2017, Ono2018, Harikane2022}. Cosmological hydrodynamical simulations within the standard \(\Lambda\)CDM framework reproduce the observed UVLF over $z\le10$ \citep[][]{Behroozi2019, Vogelsberger2020, Vijayan2021, Bird2022, Kannan2022THESAN, Kannan2022MTNG, Kanna2025}, but require dust attenuation to match the shape of the UVLF in the high luminosity end. The evolution of the high-luminosity end of the UVLF therefore provides stringent constraints on dust formation and evolution models. The validity of the dust implementations in these simulations can be tested by comparing the luminosity functions measured at other wavelengths \citep[like rest-frame B-band luminosity function provided by][]{Leethochawalit2026} predicted by these simulations.

Independent constraints on dust in the early Universe are emerging from millimetre observations with the \textit{Atacama Large Millimeter/submillimeter Array} (ALMA) \citep[e.g.][]{Dayal2022, Palla2024, Traina2024, Narayanan2025, Narayanan2026}. In addition, measurements of the UV continuum slope, \(\beta_{\mathrm{UV}}\), provide joint constraints on stellar population ages and dust content \citep[e.g.][]{Bouwens2014}. As different simulations adopt distinct dust prescriptions and attenuation curves, they predict significantly different millimetre emission and \(\beta_{\mathrm{UV}}\) distributions \citep[i.e Infrared excess (IRX) vs \buv\ relationship, e.g][]{Bowler2024}. Models calibrated to reproduce the UVLF can therefore be discriminated through comparison with these independent observables.

Deep observations with the \textit{James Webb Space Telescope} (JWST) are beginning to detect galaxies at \( z \geq 10 \), with a few systems now spectroscopically confirmed at \( z \sim 12 \) \citep{Carniani2024, Naidu2026}. Early results  indicated an excess of bright galaxies relative to extrapolations of models calibrated at low redshift \citep{Finkelstein2022, Naidu2022, Adams2024, Harikane2024, Harikane2025, Harikane2022b}. The inferred stellar masses and dust contents suggest more evolved systems than predicted by simulations tuned to match low-\(z\) observations. Spectroscopic follow-up has confirmed the redshifts of many candidates and ruled out only  a subset of low-redshift interlopers \citep{Haro2023b, Buat2023, Meyer2024}. Although continued observational progress has alleviated some of the tension \citep[e.g.][]{Weibel2025}, discrepancies persist, particularly at the highest redshifts \citep[see, e.g., UVLF estimates at \( z > 15 \) based on photometric redshifts][]{PerezGonz25}.

Recent simulations are beginning to reproduce the observed galaxy population out to \( z \sim 12 \) \citep[see][for the performance of latest set of zoom-in models]{KIMAGORA2025}, typically requiring lower levels of dust attenuation. However, they all tend to underestimate the UVLF at $z>12$. Significant uncertainties remain, including the treatment of sub-grid physics, the choice (and possible evolution) of the initial mass function (IMF), and issues related to numerical convergence and the simulation volume required to obtain statistically robust predictions.

Within the standard $\Lambda$CDM framework, semi-analytical models have attempted to reconcile these discrepancies by invoking: (i) enhanced star formation efficiencies, arising from weaker (or absent) feedback during starbursts, a reduced ultraviolet background, or stochastic star formation \citep{Sun2023,Dekel2023,Mason2023,Shen2023,Pallottini2023,Chakraborty2024, Somerville2025, Yung2025, Munoz2026}; (ii) higher light-to-mass ratios, achieved through top-heavy initial mass functions (IMFs) or the contribution of Population III stars \citep{Chon2022,Pacucci2022,Trinca2024, Ventura2024, Hutter2025}; and/or (iii) reduced dust attenuation, facilitated by spatial segregation of dust, dust ejection through radiation-driven outflows, super-Eddington accretion, or even dust-free scenarios\citep{Inayoshi2022,Ferrara2023,Ferrara2024,Ziparo2023}.
In parallel, alternative explanations beyond \(\Lambda\)CDM have also been explored. For example, the observed galaxy abundances may be reproduced by modifying the primordial power spectrum \citep[e.g. a blue tilt;][]{Hirano24, Parashari23}, introducing non-Gaussian initial conditions \citep{Biagetti23}, invoking early dark energy models \citep{Shen24, Forconi24F}, or considering extensions to \(\Lambda\)CDM and modified gravity scenarios \citep{Moffat24}.

Spectroscopic observations using JWST are very useful for constraining the stellar masses (${\rm M_\star}$), specific star formation rates (sSFRs), dust attenuation, and the average metallicity and abundance pattern in high-$z$ galaxies, in addition to providing confirmed redshifts \citep[][]{Roberts-Borsani2024, Roberts-Borsani2025, Hayes2025, Tang2026}. 
In addition, measured ratio of the Balmer emission lines \citep[In particular H$\alpha$/H$\beta$ at $\le7$, see][]{Tsujita2026, Karthikeyan2026} and \citep[$H\beta$/H$\gamma$ for even higher redshits][]{Roberts-Borsani2024}, are very useful in constraining the attenuation around the newly formed massive starts {(with age $\leq 10$ Myr)}. Measurements of emission line equivalent widths and flux ratios of different emission lines are useful in constraining the H~{\sc i} ionizing photon production efficiency and escape of such photons from high-$z$ galaxies. These are extremely important for constraining the cosmic-reionization histories. Absorption line spectroscopy using JWST will provide constraints on the spread of metals in the circumglactic medium (CGM) around galaxies and in the intergalactic medium (IGM) \citep[see for example,][]{Bordoloi2023, Zou2024}. As the efficiency of spreading metals into the CGM and IGM also depends on the parameters that govern the efficiency of star formation, any attempt made to make the high-$z$ galaxies form stars efficiently will leave its imprint in the form of  metallcities and metal covering factor in the IGM and CGM.

The \lya\ damping wing, now routinely detected in the spectra of individual galaxies over the redshift range \( 6.0 \leq z \leq 15 \) with JWST, provides independent constraints on the reionization history \citep[e.g.][]{Curtis-Lake2023, Umeda2024, Hsiao2024, Heintz2025, Mason2026}. These studies report a high incidence of strong damping wings corresponding to very large neutral hydrogen column densities (i.e. $\log\,N$(H~{\sc i}) $\geq 21.0$). For example, \citet{Heintz2025} find that the fraction of galaxies with $N$(H~{\sc i}) $>10^{21}\,\mathrm{cm}^{-2}$ increases from $\sim 60\%$ at $z \sim 6$ to $65$--$90\%$ at $z \geq 8$. Such large column densities may imply a substantially neutral intergalactic medium (IGM) surrounding these galaxies and/or a significant contribution from H~{\sc i} absorption in the ISM/CGM. Conversely, strong \lya\ emission with little or no damping wing absorption has also been reported in some \( z \sim 13 \) galaxies \citep[e.g.][]{Witstok2025}, indicating the presence of ionized bubbles around these early sources. Hydrodynamical simulations incorporating self-consistent radiative transfer, or those treating reionization in post-processing, are therefore essential for interpreting the distribution of H~{\sc i} column densities inferred from damping wing measurements \citep[see][and references therein]{Keating2024}.

The primary aim of this work is to develop a suite of cosmological hydrodynamical simulations with sufficient dynamic range and mass resolution to interpret the rapidly growing body of observational results discussed above. In particular, we explore variations in: (i) parameters governing star formation, (ii) prescriptions for feedback from supernovae (SNe) and the formation and growth of supermassive black holes (SMBHs), and (iii) models and scaling relations for dust attenuation, as well as continuum and nebular emission.

This article is arranged as follows. In section~\ref{sec:simulations}, we provide the details of our simulations and various parameters used. Light generation, dust modeling and how various observable quantities are generated are detailed in Section~\ref{sec:spec_gen}. In section~\ref{sec:results}, we calibrate the parameters used in the post-processing by fitting the UV Luminosity function over $5\le z\le 10$. We consider (i) different relationship between dust-to-metal ratio and gas phase metallicity, (ii) different extinction curves and (iii) different Initial mass functions and (iv) single and two phase models of dust distribution. Then we compare various predictions (i.e B-band luminosity function, H$\alpha$ luminosity function, relationship between UV luminosity and spectral slope, Balmer line ratio and its relationship with stellar mass and relationship between the {color excess} E(B-V) derived from the nebular lines and continuum emission) of these calibrated models with the available observations. The main summary and the discussions on our main findings are presented in Section~\ref{sec:summary}. {All magnitudes are reported in the AB system \citep{OkeGunn1983}, and the cosmology adopted throughout this work follows \citet{Plank2020} (i.e., TT,TE,EE+lowE+Lensing).}

\section{Simulation details}
\label{sec:simulations}
\begin{table*}
    \caption{Details of NINJA simulation boxes used in this study}
 \begin{tabular}{ccccl}
    \hline
    \hline
    Name & \multicolumn{3}{c}{Values used in} & Notes \\
         & L50N2040    &L150N2040   & L250N2040   & \\
    \hline
    ${\rm L_{Box}}$ & 50 & 150 & 250 & Simulation box size in $h^{-1}$ cMpc\\
    ${\rm N_{DM}}$  & 2040$^3$ & 2040$^3$ & 2040$^3$ & Number of dark matter particles\\
    ${\rm N_{gas}}$ & 2040$^3$ & 2040$^3$ & 2040$^3$ & Initial number of gas particles\\
    ${\rm M_{DM} }$ &  $1.09\times10^6$ & $2.93\times 10^7$ &$1.36\times10^8$ & Mass of a dark matter particle in $h^{-1} M_\odot$\\
    ${\rm M_{gas} }$ & $2.01\times10^5$ &$5.44\times10^6$ & $2.52\times10^7$ & Initial mass of a gas particle in $h^{-1} M_\odot$\\
    ${\rm z_{in}} $ & 189 & 99 & 99 & Initial redshift\\
    ${\rm z_{fi}}$  & 5 & 5 & 5 &  Final redshift\\
      $\epsilon_{\rm grav}$ & 0.82 & 2.45 & 4.08 & Gravitational softening length in  $h^{-1} \mathrm{kpc}$\\
      Seed & 140425 & 150478 & 121024 & Seed used to generate Initial Conditions\\
    \hline
 \end{tabular}   
    \label{tab:sim_vars}
\end{table*}

\begin{table*}
	\centering
	\caption{Parameters that are kept same for all models. }
    \begin{tabular}{ccl}
		\hline
            \hline
		Name & Value & \multicolumn{1}{c}{Description}\\
		\hline
        \\
            \multicolumn{3}{c}{\bf COSMOLOGICAL PARAMETERS}\\
            $h$ & 0.6736 & Hubble parameters in units of 100 km s$^{-1}$ Mpc$^{-1}$\\
            $\Omega_\Lambda$ & 0.6847 & Dark energy density\\
            $\Omega_0$ & 0.3153 & Matter density\\
            $\Omega_\text{b}$ & 0.0493 & Baryon density corresponds to baryon fraction $f_\text{b}=\Omega_\text{b}/\Omega_0\approx0.156$\\
            $T_\text{CMB}$ & 2.7255 & Present day CMB temperature in Kelvin\\
            $\sigma_8(z=0)$ & 0.8111 & Present day variance of fluctuations averaged over 8Mpc scale\\
            $A_\text{s}$ & $2.1\times10^{-9}$ & Amplitude of primodial fluctuations\\
            $n_\text{s}$ & 0.9649 & Index of primodial fluctuations\vspace{4px}\\
            \multicolumn{3}{c}{\bf STAR FORMATION}\\
            $\rho_{\star\text{c}}$ & 57.7 & Critical over density for star formation\\
            $\beta$ & 0.1 & Fraction of the gas energy which is locally returned as supernovae on star formation\\
            $T_\text{C}$ & $10^3$ & Temperature of the cold star forming clouds in K\\
            $T_\text{SN}$ & $10^8$ & Temperature of the supernova remnants in K\\
            $A_0$ & $1000$ & Parameter of the SH03 model controlling the energy of the hot gas\\
            $t_{\star 0}$ & 1.5 & Maximum star formation time in units of density threshold\\
            $N_\text{g}$ & 4 & Number of star particle produces per gas particle\vspace{4px}\\
            \multicolumn{3}{c}{\bf STELLAR WIND FEEDBACK}\\
            $\sigma_{0}$ & 353 km/s & Square root of energy ejection rate. Controls mass loading\\
            $\kappa_{\text{w}}$ & 3.7 & Wind speed to local particle velocity dispersion\\
            $l_{\text{free}}$ & 20 & Wind free travel length in kpc h$^{-1}$\\
            $t_{\text{free}}$ & 60 & Wind free travel time in Myr $h^{-1}$\\
            $\rho_{\text{free}}$ & 0.1 & Wind free travel density factor in units of threshold density\vspace{4px}\\
            \multicolumn{3}{c}{\bf HALO FINDER}\\
            $L_\text{ll}$ & 0.2 & FOF Halo linking length\\
            $N_\text{FOF}$ & 32 & FOF Minimum particle count\vspace{4px}\\
            \multicolumn{3}{c}{\bf BLACKHOLE SEEDING, ACCRETION}\\
            $M_{\text{HALO}}^{(0)}$ & $5\times 10^9 M_\odot\ h^{-1}$ & Minimum FOF halo mass for seeding.\\
            $M_{\text{STAR}}^{(0)}$ & $1\times 10^7 M_\odot\ h^{-1}$ & Minimum FOF stellar mass for seeding\\
            $M_{\text{BH,min}}^{(0)}$ & $3\times 10^4 M_\odot\ h^{-1}$ & Minimum seed blackhole mass\\
            $M_{\text{BH,max}}^{(0)}$ & $3\times 10^5 M_\odot\ h^{-1}$ & Maximum seed bloackhole mass\\
            $n_{\text{BH}}^{(0)}$ & $-2$ & Seeding power law index.\\
            $\alpha_\text{acc}$ & 100 & Accretion factor.\\
            $\beta_\text{edd}$ & 2 & Eddington Factor\vspace{4px}.\\
            \multicolumn{3}{c}{\bf BLACKHOLE FEEDBACK}\\
            $\epsilon_\text{TF}$ & 0.05 & Blackhole feedback efficiency for thermal feedback.\\
            $\chi_\text{thr,max}^0$ & 0.1 & Kinetic feedback Eddington threshold upper cap\\
            $\chi_\text{thr}^0$ & 0.002 & Kinetic feedback Eddington threshold normalization\\
            $M_\text{pivot}$ & 0.01 $=10^8 M_\odot\ h^{-1}$ & Kinetic feedback Eddington threshold pivot mass\\
            $\beta_\text{pivot}$ & 2 & Kinetic feedback Eddington threshold scaling law index\\
            $\epsilon_\text{KF,max}$ & 0.2 & Kinetic feedback efficiency upper cap\\
            $f_\text{thres}$ & 0.05 &  Kinetic feedback density thresold efficiency factor\\
            $f_\text{re}$ & 20 & Kinetic feedback release factor that regulates its burstiness\\
            \hline
	\end{tabular}
	\label{tab:sim_param}
\end{table*}

The main motivation of running the NINJA ({\bf N}ISER\footnote{\href{https://www.niser.ac.in}{https://www.niser.ac.in/}}-
{\bf I}UCAA\footnote{\href{https://www.iucaa.in/en/}{https://www.iucaa.in/en/}} 
{\bf N}ew Simulations of {\bf J}WST\footnote{\href{www.jwst.nasa.gov}{www.jwst.nasa.gov}} 
G{\bf A}laxies and Quasars) suite of simulations is to reproduce the rapidly improving observational properties of galaxies, Active Galactic Nuclei (AGN) and IGM at $z\geq5$. 
For this purpose, we use the \software{MP-GADGET} code \citep{Feng2018}, a massively parallel cosmological hydrodynamics solver also used by the \ASTRID\ simulation suite \citep{Bird2022,Ni2022}. 
We typically consider simulations with $2\times 2040^3$ particles, corresponding to $2040^3$ dark matter particles and $2040^3$ baryonic particles, in three simulation boxes with comoving side lengths of 50, 150, and 250h$^{-1}$Mpc, labeled L50N2040, L150N2040, and L250N2040, respectively. These models were run at IUCAA using their PEGASUS high performance computer (HPC).

\begin{figure}
    \centering
    \includegraphics[bb=8 15 380 412,clip=true, width=\columnwidth]{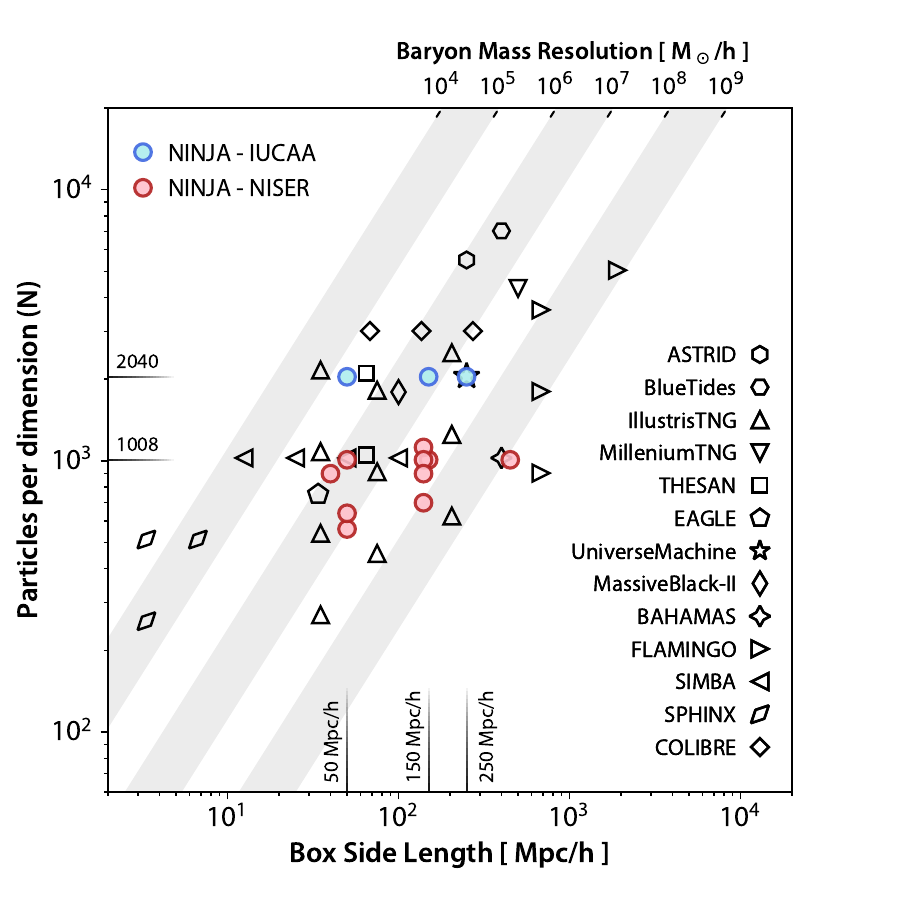}
    \caption{Resolution comparison of NINJA simulation suite with other cosmological hydrodynamical simulations. The results presented in this work correspond to the three NINJA-IUCAA boxes L50N2040, L150N2040 \& L250N2040 shown as blue filled circles. Red filled circles are NINJA runs carried out at NISER (not discussed in this paper). The diagonally shaded region denotes one decade in baryonic mass resolution. The simulations shown for comparision are from \COLIBRE\ \citep{Schaye2026COLIBRE}, \FLAMINGO\ \citep{Schaye2023}, \MilleniumTNG\ \citep{Pakmor2023}, \ASTRID\ \citep{Bird2022}, \THESAN\ \citep{Kannan2022THESAN}, \IllustrisTNG\ \citep{Nelson2019}, \UniverseMachine\ \citep{Behroozi2019}, \SIMBA\ \citep{Dave2019}, \BAHAMAS\ \citep{McCarthy2017}, \SPHINX\ \citep{Rosdahl2018SPHINX},  \BlueTides\ \citep{Feng2016}, \EAGLE\ \citep{Crain2015} and \MassiveBlackII\ \citep{Khandai2015}. }
    \label{fig:compare_ninja}
\end{figure}

In addition, we have a set of simulation boxes run at NISER. These simulations are lower in resolution than their IUCAA counterparts, but few of these simulations are evolved to $z=0$ \citep[see for example,][]{Magare2025}. Some of these runs adopt different cosmological parameters and variations in simulation-related parameters. Although the results from these simulations are not presented in this paper, we mention them for completeness of the simulation suite. The NISER simulations with matching cosmology and parameters will be incorporated into future studies. Figure~\ref{fig:compare_ninja} provides the box length vs {particles per dimension} plot for various available simulations, including NINJA runs.
The above mentioned three boxes, simulated at IUCAA, are used to cover a large range of stellar and dark matter masses of halos and galaxies. We also used the overlapping ranges in the parameter space to quantify the convergence of different physical quantities between these simulations. These are then used to present the combined results that account for the possible convergence issues, if any. The details of this procedure are presented in the Appendix~\ref{sec:res_cor}.

The configuration details of simulations used in this work are presented in Table~\ref{tab:sim_vars}. This table gives details of the mass resolution and gravitational softening length achieved in our simulation boxes. Our simulations roughly cover the length and mass scales of the \IllustrisTNG\ suite of simulations (in particular, TNG50, TNG100 and TNG300) with a slightly lower mass resolution. The L50N2040 simulation achieves a mass resolution approximately 9 times higher than \ASTRID\ (i.e 9.63$\times10^6 M_\odot$), although within a volume that is 125 times smaller. This has a mass resolution of 4.22 times less compared to TNG50 \citep{Nelson2019TNG50}. In contrast, the L250N2040 simulation spans the same volume as \ASTRID\ but at a mass resolution $\sim$14 times lower. Compared to TNG300, L250N2040 has a mass resolution 2.3 times lower \citep[see for example,][]{Springel2018}. 
Next, we briefly outline the key physical processes implemented in the simulations which closely follow those adopted in the \ASTRID\ simulations. For details of the adopted subgrid prescriptions and their implementation within \software{MP-GADGET}, we refer the reader to \cite{Bird2022} and \cite{Ni2022}. The values of parameters that govern different physical processes are summarized in Table~\ref{tab:sim_param}.

\subsection{Overview of Simulation Setup and Feedback Processes}
The power spectrum and transfer functions at the starting redshift (i.e $z_\text{in}$ in Table~\ref{tab:sim_vars}) are generated with \texttt{CLASS} \citep{Lesgourgues2011} using Cosmological parameters from \cite{Plank2020} (i.e., TT,TE,EE+lowE+Lensing). These parameters (which are slightly different from those used in ASTRID) are listed in Table~\ref{tab:sim_param}. The dark matter uses the grid, while baryons uses the glass and uses different transfer function. These are then used to set initial conditions for dark matter and baryons based on the Zeldovich approximation \citep{Zeldovich1970} with periodic conditions using \texttt{MP-GenIC}. 

Once initial conditions (ICs) are generated, gravitational forces are computed using the TreePM algorithm \citep{Xu1995,Bagla2002}, where short-range interactions are evaluated with the Tree method \citep{BarnesHut1986} and long-range interactions with the Particle-Mesh (PM) scheme. The system evolves using an N-body code with a Leap-Frog (KDK) integrator. Hydrodynamics are solved with the pressure–entropy formulation of smoothed particle hydrodynamics \citep{Hopkins2013,Reed2010,Monaghan1992}, employing a cubic density kernel \citep{Price2012}.

Unlike in the case of \ASTRID\ \citep[][]{Bird2022},  we do not model hydrogen and helium reionization (i.e patchy reionization) rather we ran our simulations using a spatially uniform but time-varying UVB (Ultra-Violet ionizing Background) given by \citet[][December 2011 update of the spectrum]{faucher2009}. In particular, we used the photoionization and heating rates for H~{\sc i}, He~{\sc i} , and He~{\sc ii} computed for this UVB. We also assumed ionization equilibrium in our calculations of the ionization state of the gas. For the optically thick cases the shielding prescribed by \citet{Rehmati2013} is implemented. The photoionization and photo-heating terms go to zero at $z>14$ for the assumed UVB. Note that \IllustrisTNG\ simulations use the same background, but the effect is instantaneously switched on at $z=6$ \citep[see for example,][]{Pillepich18}, whereas in our case a gradual buildup is considered.

The star formation model follows that of \citet{Feng2016}, adopting the multiphase ISM treatment of \citet{SpringelHernquist2003}. The chemical evolution and metal return are described in section 2.8 of \citet{Bird2022}. Radiative cooling of gas includes primordial \citep{Katz1996} and metal-line components and metallicity-dependent rates scaled according to \cite{Vogelsberger2020}. The formation of H$_2$ in low-metallicity gas and its effect on star formation are modeled following \cite{KrumholzGnedin2011}. Star formation proceeds stochastically at each timestep, with individual gas particles able to spawn up to four stellar generations before full conversion to prevent excessive metal enrichment.

For stellar wind feedback, the ejection velocity is set to be isotropic and is as given by \cite{Okamoto2010}, i.e., 
\begin{align}
v_w = \kappa_w \sigma_{\mathrm{DM}}^{1\mathrm{D}}
\end{align}
where $\sigma_{\mathrm{DM}}^{1\mathrm{D}}$ denotes the one-dimensional velocity dispersion computed from the 40 nearest dark matter particles, and $\kappa_w$ is a dimensionless parameter set to 3.7 following \cite{Vogelsberger2013}. The corresponding mass loading factor is defined as
\begin{align}
\eta_w = \left( \frac{\sigma_\text{DM}^{1D}}{353~\mathrm{km~s^{-1}}} \right)^{-2}.
\end{align}
Equations (1) and (2) do not evolve with redshift.
The wind particles are hydrodynamically decoupled and are recoupled once 60 Myr has elapsed, the wind particle has traveled a distance of 20 kpc, or the local gas density has decreased by a factor of 10 relative to its launch value. For models presented here, the wind parameters (also summarized in Table~\ref{tab:sim_param}) used are identical to that used in \ASTRID\ \citep[][]{Bird2022}.  However, the wind implementation used here differs from that used in \IllustrisTNG\ \citep{Pillepich2018TNG}. There, the wind velocity for a given ${\rm \sigma_{DM}^{1D}}$ depends on redshift and also has a minimum velocity floor, and $v_w$ is not allowed to become arbitrarily small. The mass loading factor $\eta_w$ also has a metallicity dependence. Therefore, the wind velocity and the mass loading factor assumed in our simulations will be stronger than what has been achieved for \IllustrisTNG\ at $z>6$. The effect of changing the wind parameters will be discussed in our upcoming paper.

The mass and metals returned to the ISM gas are computed for the Chabrier IMF \citep{Vogelsberger2013}. Stellar lifetimes and nucleosynthetic yields for asymptotic giant branch (AGB) stars, Type II supernovae (SNII), and Type Ia supernovae (SNIa) are adopted from precomputed tables \citep[see][for details]{Bird2022}. The code tracks the enrichment of nine chemical elements: H, He, C, N, O, Ne, Mg, Si, and Fe. Stars in the mass range $1 - 8~{\rm M_\odot}$ return mass and metals via the AGB channel, while those with masses ${\rm 8 - 40~M_\odot}$ contribute through the SNII channel. The yield tables for the Chabrier IMF are computed over the mass range $0.1 - 40~{\rm M_\odot}$, which accounts for approximately 94\% of the total IMF mass. The returned mass and metal yields are, therefore, rescaled to account for the missing fraction. Adopting an upper stellar mass limit of $300~{\rm M_\odot}$, instead of $100~{\rm M_\odot}$, results in an increase of approximately 5\% in the total returned mass and metal yields.

\begin{figure*}
    \centering
    \includegraphics[bb=20 7 845 275, clip=true, width=\textwidth]{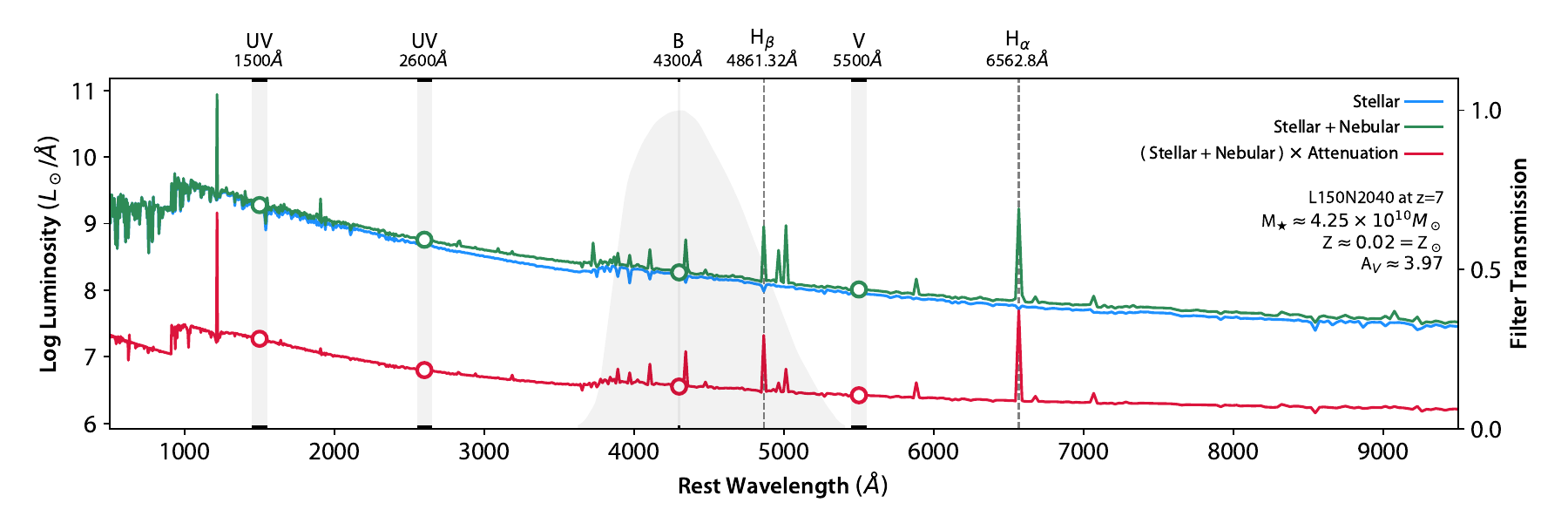}
    \caption{ A sample rest frame spectra of a $z\sim7$ halo generated by our spectral systhesis code. The gray shaded regions indicate different spectral ranges used to calculated fluxes that are compared with observational measurements.
    The UV-band (1500$\pm$50\AA) is used for UV magnitude, B-band (4300\AA) with Johnson filter (shown as shaded transmission curve) is used for B-band magnitude. Both the magnitudes are used to obtain luminosity functions to compare with observations. The $\text{H}_\alpha$ and $\text{H}_\beta$  line flux is calculated at emission line wavelength 6562.8\AA\ and 4861.32\AA\ respectively to obtain Balmer ratio $\text{H}_\alpha/\text{H}_\beta$ and $\text{H}_\alpha$ luminosity function.\ Photometry points at 1500\AA\ and 2600\AA\ with band width 100\AA\ is used to get UV slope $\beta_\text{UV}$ assuming power law $f_\lambda = \lambda^{\beta_\text{UV}}$. The Balmer ratio is used to get nebular color excess $E(B-V)_\text{nebular}$. For stellar color excess $E(B-V)_\text{stellar}$ photometry points at B-band (4300\AA) and V-band (5500\AA) is used with band width 20\AA\ and 100\AA\ respectively. A smaller band width is used in B-band to avoid line contamination from $\text{H}_\gamma$.
    }
    \label{fig:rest_frame_sed}
\end{figure*}
Black holes are seeded in halos (identified on the fly using the Friends-Of-Friends, FOF) with  masses  greater than $5\times10^9 {\rm M_\odot} h^{-1}$ and stellar masses exceeding $10^7 {\rm M_\odot} h^{-1}$, ensuring adequate stellar particle resolution for dynamical friction to operate effectively and allow the black hole to settle toward the halo center. For halos satisfying these conditions, seed black holes are drawn from a power-law mass distribution with an index of –2, spanning $3\times10^4 {\rm M_\odot} h^{-1}$ to $3\times10^5 {\rm M_\odot} h^{-1}$, representing a population of low-mass black hole progenitors. Subsequent growth occurs through gas accretion, modeled using a Bondi–Hoyle–Lyttleton–like prescription \citep[Refer to][for details.]{Ni2022}. The rate is capped at the Eddington limit, and both the Bondi and Eddington terms are scaled by a subgrid correction factor to account for unresolved density and temperature structures below the spatial resolution of the simulation, i.e.,
\begin{align}
    \dot{M}_\mathrm{BH}=\mathrm{min}(\alpha_\mathrm{acc} \dot{M}_\mathrm{BHL},\beta_\mathrm{Edd} \dot{M}_\mathrm{Edd}).
\end{align}
In our case, we use $\alpha_{\rm acc} = 100$ and $\beta_{\rm Edd} = 2$.
The supermassive black hole (SMBH) is assumed to radiate with a bolometric luminosity $L_\mathrm{bol}$ that is proportional to its accretion rate. This relationship is expressed as:
\begin{align}
    L_\mathrm{bol} = \eta_{BH} \dot{M}_\mathrm{BH}c^2
\end{align}
where $\eta_\mathrm{BH}$ denotes the mass-to-light conversion efficiency in the accretion disk, adopted from the standard thin-disk model of \cite{ShakuraSunyaev1973} and is set to a fixed value of 0.1. A fraction of 5\% of this emitted luminosity is coupled to the surrounding gas and converted into thermal energy, regulating thermal feedback processes in the host galaxy. The feedback switches to kinetic mode when the accretion rate drops below the threshold Eddington ratio, i.e. $\dot{M}_\text{BH}/\dot{M}_\text{Edd}<\chi_\text{thr}$ where the threshold scales with black hole mass as \citep[see][]{Weinberger2017}
\begin{align}
    \chi_\text{thr} &= \text{min}\left[{\chi_\text{thr}^0}\left(\frac{M_\text{BH}}{M_\text{pivot}}\right)^{\beta_\text{pivot}},\chi_\text{thr,max}^0\right].
\end{align}
Here, $M_{\rm pivot}$ and $\chi_{\rm thr}^{0}$ set the characteristic mass scale and normalization of the transition threshold, respectively, and jointly determine the black hole mass at which kinetic feedback becomes important. A positive power-law index, $\beta_{\rm pivot}>0$, causes the threshold Eddington ratio to increase with black hole mass, thereby strongly suppressing low-mass black holes from entering the kinetic mode and keeping them predominantly in the thermal mode. The upper cap, $\chi_{\rm thr,max}^{0}$, limits the threshold at high masses such that sufficiently massive black holes accreting at low Eddington ratios preferentially transition to kinetic feedback. Consequently, low-mass haloes are only weakly affected by kinetic feedback, whereas massive galaxies are efficiently quenched once their central black holes grow sufficiently massive and enter the low-accretion kinetic regime. In this mode, the feedback energy is given by $\Delta \dot{E}_\text{KF}= \epsilon_\text{KF} \dot{M}_\mathrm{BH}c^2 $ where the efficiency is
\begin{align}
    \epsilon_\text{KF} = \text{min} \left( \frac{\rho}{f_\text{thres}\rho_\text{SF,thres}} , \epsilon_\text{KF,max}  \right).
\end{align}
Here, $\rho_\text{SF,thres}$ is the density threshold for star formation (i.e $\rho_{\star c}$ in Table~\ref{tab:sim_param}),  $\epsilon_\text{KF, max}$ is the kinetic feedback efficiency cap of $20\%$ 
and $f_\text{thres}$ is  its coupling efficiency factor.
The energy is accumulated over time and released once the accumulated kinetic feedback energy exceeds
\begin{align}
    E_\text{inj,min} =  \frac{1}{2}  \sigma_\text{DM}^2 m_\text{enc} f_\text{re}.
\end{align}
Here, $\sigma_{\rm DM}$ is the one-dimensional dark matter velocity dispersion, $m_{\rm enc}$ is the gas mass enclosed within the feedback region, and $f_{\rm re}$ is the release factor that regulates the burstiness of the kinetic feedback.
The main differences with respect to \ASTRID\ are that the seeding condition used here requires five times more stellar mass, and the power law slope adopted for black hole mass sampling is steeper (power law index used is $-2$ in NINJA as compared to $-1$ in ASTRID).

\section{Spectral energy distribution of galaxies}
\label{sec:spec_gen}

\subsection{Stellar and diffuse continuum emission}
\label{sec:sed}
For each identified halos (using FOF) we generate the spectrum as a sum of stellar continuum and nebular (continuum + line) emission from the H~{\sc ii} regions surrounding individual star particle.
We obtained the intrinsic stellar continuum emission for each star particle using 
the simple stellar population (SSP) synthesis  model {\sc bpass}-v2.2.1 \citep{Eldridge2017,Stanway2018} based on its mass, age [51 grid points in the range $6\le \log_{10}(\tau(\text{yrs})) \le 11$] and metallicity [13 grid points in the range $-3.30\le \log_{10} [Z/Z_\odot]\le+0.30$]. 
{The {\sc bpass} templates are available for an instantaneous starburst of $10^6~{\rm M_\odot}$.}
{Accounting for binary interactions in {\sc bpass} results in bluer SEDs at older ages and significantly enhances the production of ionizing photons.} For each of these templates, we use \software{CLOUDY} \citep[version C23;][]{Cloudy_C23}  to generate the nebular emission as a sub-grid contribution for individual star particle. The gas cloud is assumed to have a spherical geometry with an inner radius of 1 pc and a constant hydrogen number density of ${\rm n_H}=100~\mathrm{cm}^{-3}$ \citep[see for example,][]{Byler2017}.
Elemental abundance pattern is assumed to follow the solar relative abundances following \cite{Grevesse2010}, 
with all absolute abundances scaled to match the metallicity used to generate the {\sc bpass} {template} spectra. \software{CLOUDY} produces both the nebular continuum and emission line spectra. The simulation is stopped when the gas temperature decreases below 100~K or when the ionization fraction falls below 1\%. Note that for models discussed here we do not include the effect of dust in the H~{\sc ii} regions. 

In this work, we mainly focus on the contribution of nebular emission to the continuum flux. We note that UV luminosity is not very sensitive to the inner radius (varied between 1 and 10 pc) for a given density and metallicity of the gas. On the other hand, for a given inner radius, the UV luminosity varies within 5\% when the density is varied between 1 to 10$^{-5}$ cm$^{-3}$. We note that the H$\alpha$ line luminosity also does not vary appreciably over the density range considered here. We plan to have a more thorough exploration of parameters in \software{CLOUDY} models in the upcoming paper when we look at nebular emission lines in detail. We assume a complete covering factor for the H~{\sc ii} region producing nebular emission. Allowing ionizing UV radiation to escape (i.e., the escape fraction $f_\text{esc}<1$) will reduce diffuse emission. 

For each star particle, we generate the spectrum (i.e both stellar and nebular components) using the nearest template in terms of its metallicity and age (calculated from its birth time).
Then the {\sc bpass} templates as well as the nebula emission spectrum from \software{CLOUDY} are scaled linearly by the mass of star particle. Such scaled SEDs for different stars in a galaxy are added to represent the composite spectrum ($F_{\lambda}$) of the galaxy, 
\begin{align}
    F_{\lambda}(z)=\sum_{\text{i}=1}^{\text{i}=N} \left(\frac{m_{i}}{10^6M_\odot}\right) [f_{\lambda}^{\text{st}}\left(\tau_i(z),Z_{i}^{\text{birth}}\right) + f_{\lambda}^{\text{nb}} \left(\tau_i(z),Z_{i}^{\text{birth}}\right)] .
\end{align}
Here, $f_{\lambda}^{\rm st}$  and $f_{\lambda}^{\rm nb}$ are the contributions of the stellar and nebular emission obtained for 
an $i$th star particle ({out of total $N$ star particles in the halo}) of mass $m_i$ and age $\tau_i(z)=t_i(z)-t_{i}^{\text{birth}}$ where $t_i(z)$ is the time of snapshot (corresponding to age of the Universe in the given cosmology) at redshift $z$.
\ The $t_{i}^{\text{birth}}$ and $Z_{i}^{\text{birth}}$ are the birth time and the inherited birth metallicity, respectively.

It is well known that the luminosity emanating from the star particle at different wavelength will be sensitive to the assumed initial mass function \citep[IMF, see for example,][]{Samui2007,Wilkins2016}.
For most of the results presented here, we use Chabrier IMF \citep{Chabrier2003} with an upper mass cut of $300~{\rm M_\odot}$ for light synthesis.
The choice of Chabrier IMF is consistent with the feedback implementation in \texttt{MP-GADGET}.
However, \texttt{MP-GADGET} adopts a stellar mass range of $0.1~{\rm M_\odot}$ to $40~{\rm M_\odot}$ for the Chabrier IMF (see Section \ref{sec:simulations}). As discussed above, extending the upper mass limit to $300~{\rm M_\odot}$ in our light synthesis, instead of the typical $100~{\rm M_\odot}$, would introduce an additional $\sim5\%$ enhancement in the total returned mass and metal yields, which are not consistently incorporated in  the simulation.

We also note that this is the chosen IMF in the case of \IllustrisTNG\ \citep{Vogelsberger2020}, \FLARES\ \citep{Vijayan2021},
and \THESAN\ \citep{Garaldi2024} simulations. Although these simulations consider the  mass of stars in the range 0.1-100 M$_\odot$, 
we tend to produce more UV luminosity for a given star formation rate because our upper limit being 300 ${\rm M_\odot}$.
We examine the impact of IMF variations on the observed UV luminosity function in Section~\ref{sec:UVLF}. In particular, we discuss the top-heavy Kroupa IMF with an upper mass cutoff of ${\rm 300~M_\odot}$ (hereafter referred to as KP300-TH), which yields the maximum luminosity among the available IMF models in {\sc bpass}. In this top-heavy IMF, the high mass end power-law slope is set to $-2.0$ above ${\rm 0.5~M_\odot}$, compared to $-2.3$ above ${\rm 1~ M_\odot}$ for the Chabrier IMF, thereby enhancing the relative contribution of massive stars.

\subsection{Modeling  the dust attenuation}
\label{sec:dust}
In NINJA we model the dust extinction as a post processing step.
For this we mainly use the known scaling relationship between 
hydrogen column density (both neutral as well as total) and visual extinction $A_V$ \citep[see for example,][]{Bohlin1978, Rachford2009, Gorenstein1975, Guver2009, Zhu2017, Li2024}.
Typically, $N$(H) = $\kappa A_V$, with $\kappa$ values in the range of 2.21 to 3.32 $\times 10^{21}$ cm$^{-2}$ mag$^{-1}$ \citep[see Table 1 of][]{Li2024} for the Milky Way sightlines.  In the case of LMC, SMC-wing and SMC-bar the average values of $\kappa$ are 3.25, 7.40 and 13.18 (in units of  $10^{21}$ cm$^{-2}$ mag$^{-1}$)   respectively \citep[see Table 2 of][]{Gordon2003}. In this study, we consider $\kappa = 2.2\times 10^{21}$ cm$^{-2}$ mag$^{-1}$ and use
\begin{equation}
   A_V =  \frac{N{\rm (H)} f(Z_g)}{\kappa \epsilon} 
    \label{eq:empirical_AV_to_NH}
\end{equation}
where $\epsilon$ and $f(Z_g)$ are introduced to allow different dust formation efficiency and dust-to-gas ratio (DTG) as a function of gas phase metallicity $Z_g$. In certain cosmological simulations \citep[see for example,][and references therein]{Vogelsberger2020} $f(Z_g)$ is assumed to be $(Z_g/Z_\odot)^{\alpha}$ with $\alpha = 1$ 
and $Z_\odot = 0.0127$. We consider such a scaling in our fiducial model.
However, the dependence of DTG on metallicity  seems to be steeper and is more or less well fitted with a broken power law with a much steeper slope at low metallicities \citep[see][]{Remy_Ruyer_2014}. Such a dependence of DTG on metallicity is also motivated by dust evolution models that consider various formation and destruction processes \citep[][]{Asano2013,  Zhukovska2014, Galliano2021}. To capture this, we also consider the case where $f(Z_g)$ follows a double power-law with a break at $Z_g = 0.1862 Z_\odot$ having slopes $1.00$ and $3.08$ at high and low metallicity ends, respectively.  We keep $\epsilon$ as a free parameter and obtain its best fit value at different redshifts by fitting the observed UV luminosity function. Following Table~1 of \citet{Remy_Ruyer_2014}, it is taken that the functional form of the double power law is 
\begin{align}
    f(Z_g) = \begin{cases}
        Z_g/Z_\odot & Z_g < Z_\textrm{cut}\\
        (Z_g/Z_\odot)^3 \times(Z_\textrm{cut}/Z_\odot)^{-2} & Z_g \ge Z_\textrm{cut}\\
    \end{cases}
    \label{eqn:doublePL}
\end{align}
with $Z_\textrm{cut}=0.1862\ Z_\odot$ and we assume $Z_\odot=0.02$.

To get the reddened spectrum we also need to assume the wavelength dependence of attenuation  (i.e $A_\lambda$).
The $A_\lambda$ is approximated by
\begin{align}
{\rm A_\lambda} = {\rm A_V} \times \zeta(\lambda/5500{\textrm\AA})
\end{align}
where, $\zeta$ is the attenuation curve. In the local universe, a wide range of optical-to-UV slope is observed for the derived attenuation curves. There is a trend of this slope being inversely correlated with the column density of dust 
\citep[or equivalently the visual extinction $A_V$, see][]{Salim2020}.
The ever growing JWST based measurements are providing more information on attenuation curves at the redshift range of our interest (i.e $z>5$). The existing JWST measurements till date suggest the average powerlaw slope and 2175\AA\ UV-bump strengths are less at these redshift compared to the local universe \citep[][]{Markov2025}.

On the other hand \citet{Fisher2025} have found a wide range of attenuation curves among the galaxies studied by them, the slope being systematically lower than what has been seen for SMC and slightly steeper than that of \citet{Calzetti2000}. They also found $\sim$30\% of their sample to show 2175\AA\ UV bump albeit with lower strength. \citet{Shivaei2025}, with the larger sample of galaxies, have found the trend of steeper slope at lower $A_V$  even at high-$z$ albeit the slope being systematically flatter for a given $A_V$. They suggested that the attenuation curves at $7\le z\le 9$ are even shallower than that of \citet{Calzetti2000}.
Here,
we consider the attenuation curve by \citet{Calzetti2000} (without the UV-bump) as our fiducial one. However, we will also consider different extinction curves to investigate their effect on the derived parameters. {Specifically, we consider the SMC-average extinction law\footnote{ We note the distinction between extinction and attenuation: the former accounts for absorption and scattering of light out of the line of sight, whereas the latter additionally includes in-scattering and incorporates the effects of complex dust–star geometries.} \citep{Gordon2024}, which exhibits a steeper wavelength dependence compared to our fiducial Calzetti attenuation law.}

It is well documented that measured $A_V$ using the stellar continuum and Balmer decrement in galaxies at different redshifts are not consistent with each other \citep[for example,][]{CharlotFall2000}. The larger extinction observed for the nebular line emission is interpreted as photons from young stars suffering additional reddening due to dust in the parent molecular clouds (denoted as birth clouds, BC).  In principle, the contribution to ${A_V}$
comes from dust in the parent molecular clouds in which stars are formed (denoted by ${\rm A_V^{BC}}$) and from the line of sight dust in the interstellar medium (i.e dust in the gas particle in our simulation box, ${\rm A_V^{ISM}}$), i.e.
\begin{align}
A_V = A_V^{\mathrm{ISM}} + A_V^{\mathrm{BC}}
\end{align}
as suggested by \citet{CharlotFall2000}.
As our simulations do not resolve the molecular clouds we follow
\citet{Vogelsberger2020}, where  
$
A_V^{\mathrm{BC}} = 2 \langle A_V^{\mathrm{ISM}} \rangle
$  
is applied for stars whose lifetime is less than 
$10$ Myr \citep{BlitzShu1980}. We call this "birth cloud attenuation" model.
For simplicity, in both cases we assume the attenuation curve to be the same. However, in the literature, it is common practice to assume a power-law attenuation curve for the birth cloud component.
Note, on the other hand, \citet{Vijayan2021} have assumed both V-band extinction from the birth cloud and ISM to scale with local metallicity and treat the scaling factors as free parameters in their \FLARES\ simulation. {A comparison of differences in light synthesis models across various cosmological simulations is presented in  Appendix~\ref{sec:appendix-comp-other-sim}.}

\subsubsection{Visual extinction from simulations  }
First we compute $A_V$ towards each star in an identified halo for the assumed form of $f(Z_g)$.  For this we set uniformly spaced grid points, along a chosen direction ({$+$Z axis in this work}), with a spacing $\Delta s$, typically 1/20$^{\rm th}$ of the average interparticle separation in our simulation box. This number was reached after evaluating the convergence of the column density. 
{For each grid point (indexed by "$j$"), we identify gas particles within a region defined by the maximum SPH smoothing length of gas particles in the halo. These particles contribute to the density and metallicity at the grid point according to their respective masses and SPH smoothing lengths, weighted by the SPH kernel.} Then $A_V$ along the line of sight is computed using,
\begin{align}
    A_V= \frac{(1+z)^2}{\epsilon \kappa} \frac{X}{m_\text{H}} \sum_{j}^{} f(Z_{g,j})~\bigg{[}\sum_{i} m_iW_i(r,h)\bigg{]} ~\Delta s_j
    \label{eq:neutral_hyd_column}
\end{align}
with
\begin{align}
    Z_{g,j} = \frac{{\sum_i} m_i Z_i W_i}{\sum_i m_i W_i }.
\end{align}
Here, $X=0.76$ is the abundance of hydrogen, $m_{\rm H}$ is the mass of the hydrogen atom, $m_i$ is the gas mass of the "$i^{\text{th}}$" particle  and $Z_i$ the metallicity by the mass of the $i^{\text{th}}$ particle  in units of solar metallicity. 
The $(1+z)^2$ factor ensures the sum is over the physical length scales. We use a Cubic-Spline kernel $W(r,h)=\mathcal{W}(q=r/h)$ to gather the SPH contribution given by \citep{Monaghan1992}
\begin{align}
    \mathcal{W}(q)=\frac{8}{\pi h^3}
    \begin{cases}
    1-6q^2+6q^3 & \text{for } q\leq 0.5\\
    2(1-q)^3 & \text{for }0.5<q\leq1\\
    0 & \text{for } q>1\\
    \end{cases}
\end{align}
The exact value of $\epsilon$ will be obtained for each redshift by matching our predicted UV luminosity function (LF) with the observed one  as explained in section~\ref{sec:UVLF}.

Figure~\ref{fig:rest_frame_sed}, provides an example of the rest frame spectrum (stellar in blue, intrinsic (i.e stellar + nebular) green  and attenuated in red) generated for one of the halos at $z=7$. This halo has a stellar mass of $4.25 \times 10^{10} M_\odot$ and $A_V$ = 3.97 (for the best fitted $\epsilon$ as discussed below). In this case, we have used the Calzetti attenuation curve and the DTG ratio proportional to metallicity.
The H~{\sc i} absorption (i.e., both line and Lyman continuum) from the ISM and IGM is not included, {as their impact on the rest-frame UV luminosity is found to be negligible}. The gray shaded rectangular regions mark the spectral ranges used for computing various observables. The filter response of the Johnson B-band is also shown with a gray shaded region.

\subsection{Generation of observable quantities:}
Once the spectral energy distribution (SED) is generated for each galaxy (i.e stellar+nebular before and after applying the dust attenuation), we compute the average flux at $1500~\text{\AA}$ within a band range of $100~\text{\AA}$ ($1450~\text{\AA}$ to $1550~\text{\AA}$) in the rest frame (see Figure~\ref{fig:rest_frame_sed}). This flux is then converted to the magnitude scale following \cite{OkeGunn1983},
\begin{align}
    M_\mathrm{UV} = -2.5 \log_{10}(f_\nu)-48.60
\end{align}
where $f_\nu = \lambda^2 f_\lambda / c$, is the flux from the galaxy measured at the 10 pc distance, expressed in units of erg s$^{-1}$ Hz$^{-1}$ cm$^{-2}$ and evaluated at $\lambda = 1500~\text{\AA}$. 
For the B-band magnitude, we adopt the revised Johnson filters \citep[][shown with gray shaded region in Figure~\ref{fig:rest_frame_sed}]{Mann2015}, rather than averaging over a 100~\AA\ window, in order to maintain consistency with the observational methodology of \cite{Leethochawalit2026} for the B-band luminosity function.

To obtain the photometric UV slope (i.e $\beta_{UV}$), we determine the average flux at $1500~\text{\AA}$ and $2600~\text{\AA}$, each with a band range of $100~\text{\AA}$ (see Figure~\ref{fig:rest_frame_sed}), and fit a power-law to the flux–wavelength relation to estimate the slope with $f_\lambda \propto \lambda^{\beta_\text{UV}}$.
For the $H_\alpha$ and $H_\beta$ line luminosities, we extract spectral regions centered at $\lambda_{H\alpha} = 6562.80\text{\AA}$ and $\lambda_{H\beta} = 4861.32\text{\AA}$ and subtract the average continuum luminosity measured from the neighboring pixels.
The continuum subtracted spectrum is integrated to obtain the $H_\alpha$ and $H_\beta$ luminosities used in the Balmer ratio ($H_\alpha/H_\beta$) analysis.
The continuum color excess is calculated as $E(B-V)_\text{star}=(B-V)_{\text{obs}}-(B-V)_{\text{int}}$. Here attenuated SED is taken as observed SED and stellar with nebular contribution is take as intrinsic SED. The nebular color excess is calculated from the Balmer ratio as
\footnote{
{In the literature, this relation is often expressed using the definition $\kappa(\lambda) = A(\lambda)/E(B-V)$. In this work, however, we use the form of the attenuation-law $\zeta(\lambda) = A(\lambda)/A_V$. The two definitions are related through $R_V = A_V/E(B-V)$, such that $\kappa(\lambda) = R_V \, \zeta(\lambda)$.
}
}
\begin{align}
    E(B-V)_\text{neb} = \frac{2.5}{R_V[\zeta(H_\beta)-\zeta(H_\alpha)]}\log_{10}\left(\frac{(H_\alpha/H_\beta)_\text{obs}}{(H_\alpha/H_\beta)_\text{int}}\right).
\end{align}
For $E(B-V)$ measured from the continuum (denoted as $E(B-V)_\text{star}$) we use measured fluxes over a 20 \AA\ window centered at 4300 \AA\ and 100 \AA\ window around 5500 \AA. For the Calzetti attenuation law we use $R_V = 4.05$ \citep{Calzetti2000} with $\zeta(H_\alpha)=0.822$ and $\zeta(H_\beta)=1.136$. For the SMC extinction law we use $R_V = 3.02$ \citep{Gordon2024} with $\zeta(H_\alpha)=0.718$ and $\zeta(H_\beta)=1.213$.

\section{Results}
\label{sec:results}
\subsection{Luminosity Function}
\label{sec:UVLF}
\begin{figure*}
    \centering
    \includegraphics[width=\textwidth]{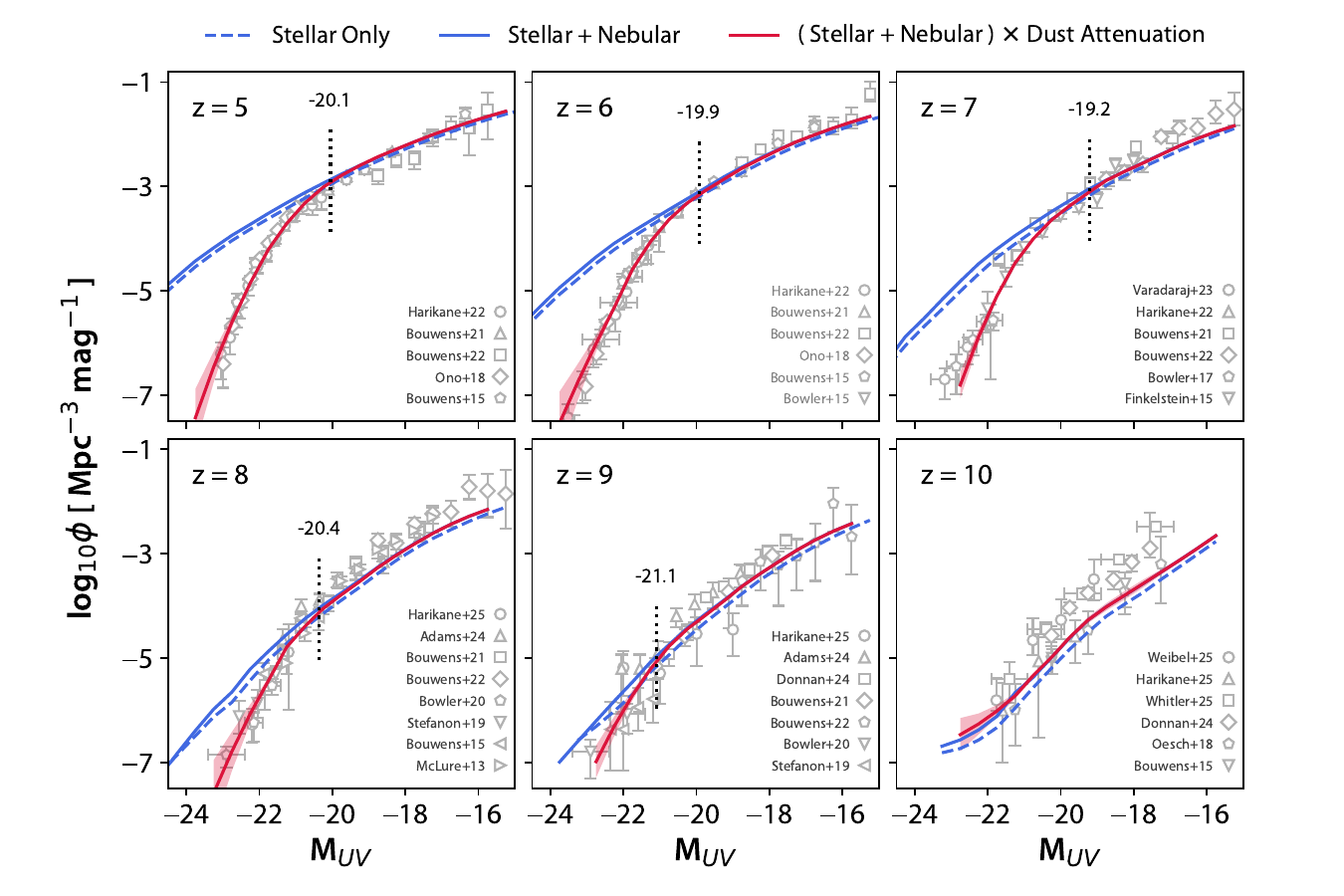}
    \caption{ 
    The observed UVLF over $5\le z\le 10$ are compared with those
     constructed by combining results of the NINJA simulations L50N2040, L150N2040, and L250N2040. Luminosities are computed at $1500\text{\AA}$ over a band of width $100\text{\AA}$. The blue solid line shows the sum of the intrinsic stellar SED (blue dashed curve) and the nebular continuum, while the red solid line includes fiducial dust attenuation applied to this composite SED. The faded red band shows the error associated with dust attenuated UVLF ({ see section~\ref{sec:UVLF} for details}).
     The vertical dotted black line in each panel shows the UV magnitude where the UVLF obtained including reddening corrections deviates from intrinsic LF by two percent (in logarithmic scale). Observational data used to minimize the $\chi^2$ are from \citet{Harikane2025, Harikane2022, Whitler2025, Weibel2025, Adams2024, Donnan2024, Varadaraj2023, Bouwens2022, Bouwens2021, Bouwens2015, Bowler2020, Bowler2017, Bowler2015, Stefanon2019,Oesch2018, Ono2018, Finkelstein2015} and \citet{McLure2013}.}
    \label{fig:ninja_LF}
\end{figure*}

\begin{table*}
    \centering
    \renewcommand{\arraystretch}{1.4}
    \begin{tabular}{lcccccc}
         \hline
         Model & $z=5$ & $z=6$ & $z=7$ & $z=8$ & $z=9$ & $z=10$  \\
         \hline
         Fiducial &$2.71_{-0.08}^{+0.09}$&$2.86_{-0.12}^{+0.13}$ & $1.64_{-0.13}^{+0.13}$ & $4.10_{-0.36}^{+0.39}$ & $10.00_{-3.16}^{+3.19}$ & ....\\
         Broken Power Scaling & $1.41_{-0.06}^{+0.06}$ & $1.26_{-0.09}^{+0.10}$ & $0.50_{-0.05}^{+0.06}$ & $1.50_{-0.17}^{+1.18}$ & $5.75_{-3.13}^{+4.37}$ & $7.00_{-3.53}^{+\infty}$\\
         SMC Extinction Law & $4.73_{-0.15}^{+0.15}$ & $4.98_{-0.22}^{+0.22}$ & $3.09_{-0.24}^{+0.25}$ & $6.21_{-0.52}^{+0.57}$ & $15.00_{-2.10}^{+7.37}$ & .... \\ 
         Birth Cloud Attenuation & $9.39_{-0.43}^{+0.43}$ & $ 9.71_{-0.34}^{+0.36}$ & $5.19_{-0.35}^{+0.38}$ & $15.40_{-1.75}^{+1.35}$ & $20.00_{-2.65}^{+15.44}$ & .... \\
         KP300-TH IMF Single & $1.03_{-0.03}^{+0.03}$ & $0.99_{-0.03}^{+0.03}$ & $0.40_{-0.03}^{+0.01}$ & $0.78_{-0.05}^{+0.05}$ & $0.81_{-0.12}^{+0.16}$ & $1.40_{-0.38}^{+0.64}$\\
         KP300-TH IMF Dual ($Z_c=0.01$) & $2.14_{-0.08}^{+0.08}$  & $2.01_{-0.09}^{+0.09}$  & $0.94_{-0.07}^{+0.07}$  & $2.36_{-0.20}^{+0.22}$ &$3.50_{-0.16}^{+1.02}$ & .... \\ 
         KP300-TH IMF Dual ($Z_c=0.001$)& $2.57_{-0.11}^{+0.11}$  & $2.81_{-0.12}^{+0.12}$ & $1.58_{-0.12}^{+0.13}$ &  $4.01_{-0.35}^{+0.37}$ & $9.00_{-2.89}^{+1.53}$ & ....\\
         \hline
    \end{tabular}
    \caption{Summary of the best-fitting $\epsilon$ values at different redshifts for various dust and IMF implementations. Details of the models are provided in Section~\ref{sec:spec_gen}. The corresponding UV luminosity functions for all models are shown in Figure~\ref{fig:ninja_lf_comp}.}
    \label{tab:lf_fit_results}
\end{table*}

\begin{figure*}
    \centering
    \includegraphics[width=\textwidth]{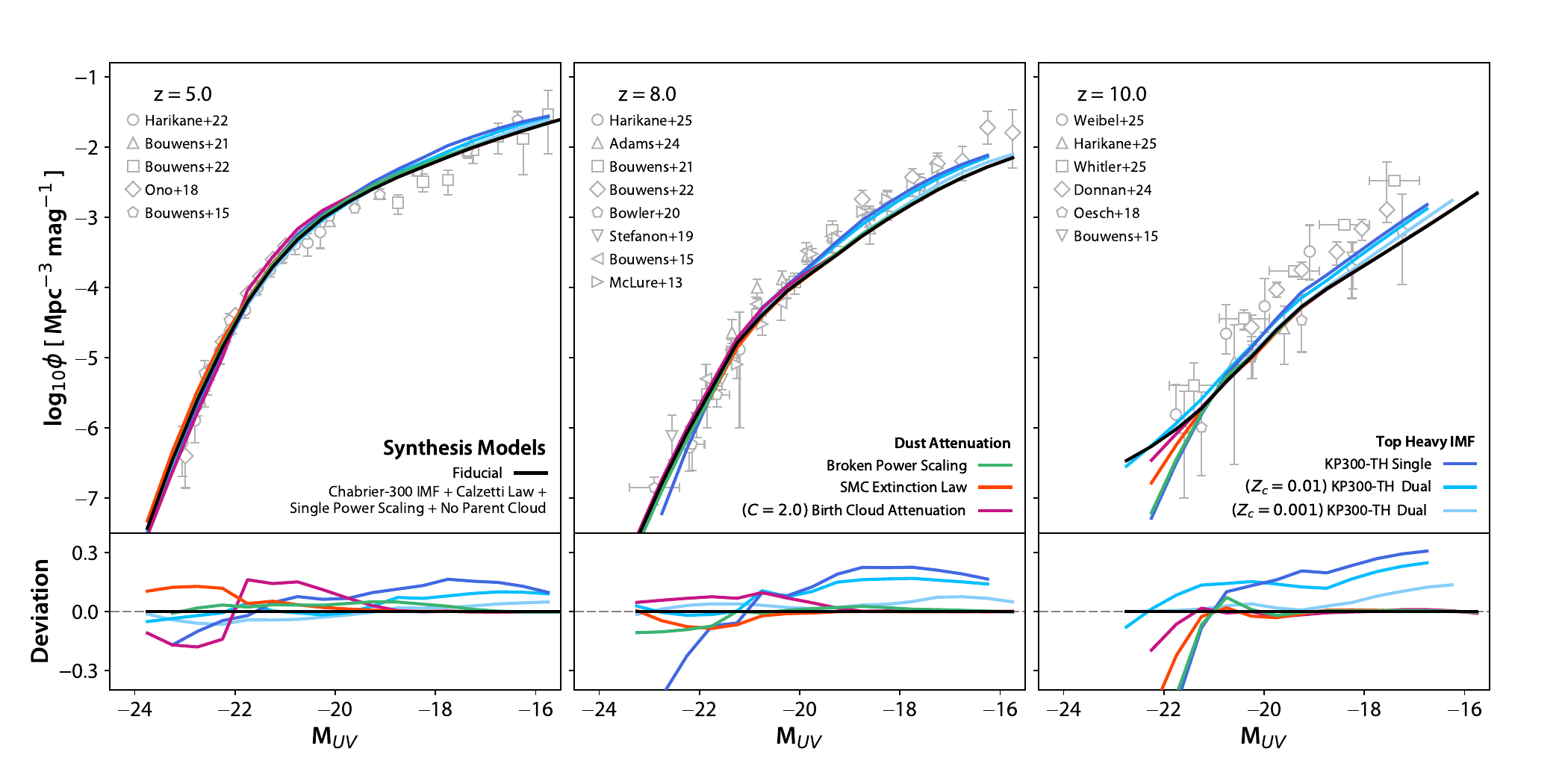}
    \caption{
    Best fits to the  observed UVLFs at three different redshifts using NINJA  simulations under different implementations of dust reddening (see section~\ref{sec:dust}). 
    Our fiducial model (black - also see Figure~\ref{fig:ninja_LF}) uses a Chabrier stellar IMF \citep{Chabrier2003} with the Calzetti attenuation law \citep{Calzetti2000}, assumes the dust column density to scale linearly with metallicity.
    The `Broken Power Scaling' model (green) adopts the dependence of dust column density on metallicity to broken power-law \citep[as provide by][]{Remy_Ruyer_2014} and rest as in our fiducial model. The `SMC Extinction Law' model (orange) uses the empirical SMC average extinction law \citep{Gordon2024} and rest as in our fiducial model. The `Birth Cloud Attenuation' model (violet) adds a birth-cloud attenuation component on top of the ISM with $A_V^{\mathrm{BC}} = 2 \langle A_V^{\mathrm{ISM}} \rangle$ to our fiducial model.
    The remaining models discussed here use different IMFs instead of Chabrier IMF used in our fiducial model. 
    The `KP300-TH Single' model (dark blue) uses a Kroupa top-heavy IMF for all stellar particles. The `KP300-TH Dual' models apply this IMF only when the stellar metallicity satisfies $Z_c = Z/Z_\odot < 0.01$ (blue) and $Z_c = Z/Z_\odot < 0.001$ (light blue) and uses Chabrier IMF for the rest. In all cases,
    the best fitted curves are obtained by minimizing $\chi^2$. The lower panels show the deviation from the fiducial model (in log scale), and as expected the faint end is affected primarily by IMF variations and not on the parameters of the dust models. 
    }
    \label{fig:ninja_lf_comp}
\end{figure*}

As in \citet{Vogelsberger2020}, we calibrate the parameters of the dust model (mainly $\epsilon$) for any given redshift by fitting the observed UV luminosity function (UVLF). For a given redshift bin we use all the available observed luminosity functions.
The measured UVLF are binned over the range \mab$= -24$ to $-15$ mag with a bin width of 0.5 mag. 
For each luminosity bin,
we generate 10,000 random realizations drawn from a normal distribution with the mean and standard deviation equal to the mean and error (in logarithmic units) of all the observed data in that bin. The mean and error ($\sigma_\text{obs}$)  of these random realizations are used as representative values in each $M_{\rm UV}$ bin. For each luminosity bin, we estimate the uncertainty in the simulated luminosity function using spatial jackknife resampling, denoted by $\sigma_\text{jk}$, based on 27 realizations. The box is partitioned into 
$3\times 3\times 3$ sub-volumes, obtained by dividing each spatial dimension into three equal segments. Each realization is constructed by excluding one sub-volume from the full simulation box.
However, this method becomes unreliable at the bright end, where the number of identified halos in our simulation box becomes small ($N <10$). In such cases, we also compute the $1\sigma$ Poisson confidence intervals $(\sigma_\mathrm{pci})$ for both sides of the distribution. These Poisson errors are symmetrized by averaging the upper and lower uncertainties. For each bin, we  then adopt the larger of the jackknife and Poisson errors as the final simulation uncertainty $\sigma_\mathrm{sim}=\mathrm{max}(\sigma_\mathrm{jk},\sigma_\mathrm{pci})$.

We then obtained the best fitted $\epsilon$ by $\chi^2$ minimization. Here, $\chi^2$ is defined as
\begin{align}
    \chi^2 = \frac{1}{N-1}\sum_{i=1}^{N}\frac{(\Phi({\rm obs}) - \Phi({\rm sim}))^2}{{\sigma_{\mathrm{obs,i}}^2
+ \sigma_{\mathrm{sim,i}}^2}}
\end{align}
Where $\Phi({\rm obs}$) and $\Phi({\rm sim})$ 
are the observed and simulated UV luminosity functions, respectively, and $i$ goes through all the bins where we have data points from both observation and simulation.

In Figure \ref{fig:ninja_LF}, we compare the observed  UVLF at $5\le z \le10$ with our model UVLF obtained by 
stitching the UVLFs from three simulation boxes (see Appendix~\ref{sec:res_cor}).  Sources for the observational points used are provided in the figure caption. 
These observed UVLFs are mainly based on photometric redshifts and therefore averaged over wider redshift bins (that differ for different authors). In each panel, we show the dust uncorrected UVLF considering only the stellar continuum (blue dashed line) and  sum of stellar and nebular continuum (blue solid line; refer to section~\ref{sec:sed} for details). It is evident that for NINJA simulations presented here, the addition of a nebular continuum increases the UVLF at all redshifts, suggesting the importance of contributions from recently formed young stars \citep[see also the discussions presented in][]{Topping2022}.

The red curves are the best fitted UVLF (considering both stellar and nebular continuum) after accounting for the dust reddening by our fiducial reddening model (i.e., the dust-to-metal ratio scales inversely with metallicity and the Calzetti extinction curve). The shaded regions provide uncertainties in the predicted UVLF from our simulations (i.e $\sigma_{\rm sim}$ defined above). For the dust prescription used here, reddening effects are found to be important at the high luminosity end (delimited by dotted vertical lines in Figure~\ref{fig:ninja_LF}). Therefore, at low luminosity side major modifications to UVLF can be achieved either by changing the parameters related to star formation and associated feedback processes used in our simulations (i.e running a set of simulations with various input parameters given in Table~\ref{tab:sim_param}) or by changing the IMF. In this work, we explore only the latter possibility, while the first possibility  will be explored in our upcoming paper.

Our best fitted UVLFs reproduce the observed UVLF well over the full luminosity range for $z$ = 5 and 6. However, we find a slight underproduction at the low luminosity end for $z\ge7$. At all redshifts, the UVLF computed considering the stellar and nebular continuum is closer to the observations compared to simulations considering only the stellar light.  Therefore, we consider the SED generated including the nebular continuum as the intrinsic spectrum of individual galaxies in all of our analysis.

 The value of $\epsilon$ needed to get the best fit UVLF at each redshift is summarized in Table~\ref{tab:lf_fit_results}.
 For \( z=10 \), several cases do not provide stringent constraints on \( \epsilon\), i.e., there is no statistically significant requirement for dust extinction. In such cases, we do not report the corresponding \( \epsilon \) values.
 The best fit $\epsilon$ we obtain at any redshift is sensitive to the observational data used in $\chi^2$-minimization. 
 In particular, effects such as (i) cosmic variance, (ii) AGN contamination, (iii)
 candidate galaxy selection based on photometry, (iv) how well incompleteness of the survey was obtained, etc., will provide large scatter in the UVLF especially in the high luminosity end and $z\ge7$. The $\epsilon$ values quoted in Table~\ref{tab:lf_fit_results} do not include the systematic effect due to the scatter in UVLF measured by different groups used here.
  For a given metallicity, $\epsilon$ decides the hydrogen column density required to produce a given $A_V$ relative to the local Milky Way relation (see equation~\ref{eq:empirical_AV_to_NH}).
  Changing $\epsilon$ is analogous to changing the dust-to-metal ratio (or dust formation efficiency) {relative to local values}. 
A large value of $\epsilon$  implies a low dust-to-metal ratio (or dust formation efficiency) at a given metallicity compared to local values. 
 We find $\epsilon$ to increase with increasing redshift over the redshift range considered here. 
 For example, in our fiducial model, we require the dust-to-metal ratio to be  $\sim$35\% of that in our galaxy to get the best fit UVLF at $z$ = 5 or 6. This becomes $\sim$25\% for $z\sim8$ and $\le10$\% for $z\ge9$. Next we explore how our requirements on dust depend on the choice of the extinction curve and relationship between metallicity and dust-to-metal ratio.

\subsubsection{UV luminosity function: Dust parameter variations}
\label{sec:lf_dust_var}
Here, we explore the effects of varying the dust models employed (as described in section~\ref{sec:dust}). The best fitted UVLFs for different cases are summarized in Figure~\ref{fig:ninja_lf_comp} and the corresponding values of $\epsilon$ are summarized in Table~\ref{tab:lf_fit_results}.  As expected, we get good fits to the UVLF for all the cases with the UVLF at the low luminosity end being nearly identical for models using the Chabrier IMF. 
It is also clear from Table~\ref{tab:lf_fit_results} that the $\epsilon$ values required to get the best fit UVLF at a given $z$ and IMF, vary by more than a factor 6 between different models. 

The highest value of $\epsilon$ is obtained when we consider the two component (i.e., BC and ISM) dust model. In this case considerable reddening occurs in the unresolved component introduced to mimic the  parent  cloud reddening, hence the general ISM is systematically less enriched with dust particles compared to our fiducial model. Note that the two component dust model is  introduced to produce differential reddening to the nebular emission lines and continuum as seen in low-$z$ galaxies (see discussions in section~\ref{sec:dust}).
This case is similar to "model-B" discussed in \citet{Vogelsberger2020} and the redshift evolution of their $\tau_{\rm dust}$ is in line with what we find for $\epsilon$ but with a steeper slope (refer to their table 3 and figure 4).  This difference could arise from differences in the simulations and/or from the UVLF data used to constrain $\epsilon$. 
{The requirement for a two-component dust model will be further discussed in Section~\ref{sec:Bal_ratio}, in the context of the Balmer line ratios.}

On the other hand, the lowest values of $\epsilon$ were obtained for models in which the dependence of the dust-to-metal ratio on the metallicity of the gas phase follows a double power law (i.e equation~\ref{eqn:doublePL}). This case requires the dust formation efficiency in the high metallicity regions to be as high as what we see locally even at $z$ = 6-8. As our models have  a correlation between stellar mass and metallicity of galaxies, we expect this model to produce distinctly different  stellar mass-dust-mass relation compared to rest of the models considered here. 

When we use the SMC extinction curve, instead of Calzetti attenuation curve, and assume the  dust-to-metal ratio to be  proportional to metallicity (i.e., our default assumption), the best fitted $\epsilon$ is $\sim$1.7 times more (i.e., the required dust-to-metal ratio is less by a factor $\sim$1.7). 
This is because the SMC extinction curve is much steeper than the Calzetti attenuation curve in the UV-optical regime. Moreover, the slope of the spectral energy distribution in the UV-optical range in this case will be different from our fiducial model, even though the fit to the UVLF is nearly identical in both cases. Therefore, $\beta_{\rm UV}$, the Balmer ratios and correlations related to them are useful in distinguishing between different extinction curves.

\subsubsection{UV luminosity function: IMF variations}
\label{sec:lf_imf_var}

Until now, we have mostly considered the Chabrier IMF for our spectral synthesis as this is consistent with the IMF used in our simulations to calculate the metal yield and SNe rates (see section~\ref{sec:sed}).
In this section, we consider possible variations to the IMF used in the spectral synthesis models (even though it is inconsistent with the one used in the simulations) to see whether we can get consistent fits to the UVLF in the low luminosity end as well. However, to keep it simple, we restricted our explorations to IMFs for which SSPs are available in {\sc bpass}.

First, we consider the Kroupa Top-Heavy IMF (KP300-TH, see section~\ref{sec:sed} for details) with an upper-mass limit extended to $300~{\rm M_\odot}$ that gives the highest UV output among the available {\sc bpass} spectral templates. The best fitted luminosity functions for this IMF for three redshifts are shown in Figure~\ref{fig:ninja_lf_comp}. The $\epsilon$ values required for the best fit are also summarized in Table~\ref{tab:lf_fit_results}. As expected, the UVLF predicted by the model using KP300-TH IMF is higher than the prediction for our fiducial case. Due to this, we need a higher dust-to-metal ratio compared to our fiducial model at all redshifts. The best fit values of $\epsilon$ suggest that a dust-to-metal ratio is of the order of or higher than what has been seen in the local universe. However, these models predict UVLF closer to the observed UVLF at the low luminosity end (for $7\le z\le9$) compared to the fiducial model. 
From Table~\ref{tab:lf_fit_results}, it is evident that the requirement of a high dust-to-metal ratio can be relaxed by using either the SMC extinction curve or the two component dust model. In both cases, the best fit values of $\epsilon$ will increase compared to what has been given in Table~\ref{tab:lf_fit_results}.

Metallicity dependent IMF is one possible way to reduce the dust optical depth at high luminosity end as there is the mass-metallicity relation.
Therefore, we consider the possibility of using the KP300-TH  IMF for low-metallicity (and therefore faint, low-mass) regions and the Chabrier IMF for the rest. We consider such a case for two cut-off metallicities {$Z_c=Z/Z_\odot$ at 0.01 and 0.001}.  The resulting best fit UVLFs are also shown in Figure~\ref{fig:ninja_lf_comp}.
These models provide a slightly better fit to the UVLF at low luminosities with realistic values of dust to metal ratio. However, the observed data points are still higher than the model predictions (see Figure~\ref{fig:ninja_lf_comp}) at low luminosities. We will address the UVLF at the low-luminosity end and its redshift evolution by changing the simulation parameters in our upcoming paper.

\subsubsection{B-band Luminosity function}
\label{sec:BLF}
\begin{figure}
    \centering
    \includegraphics[bb=18 35 405 535,clip=true,width=\columnwidth]{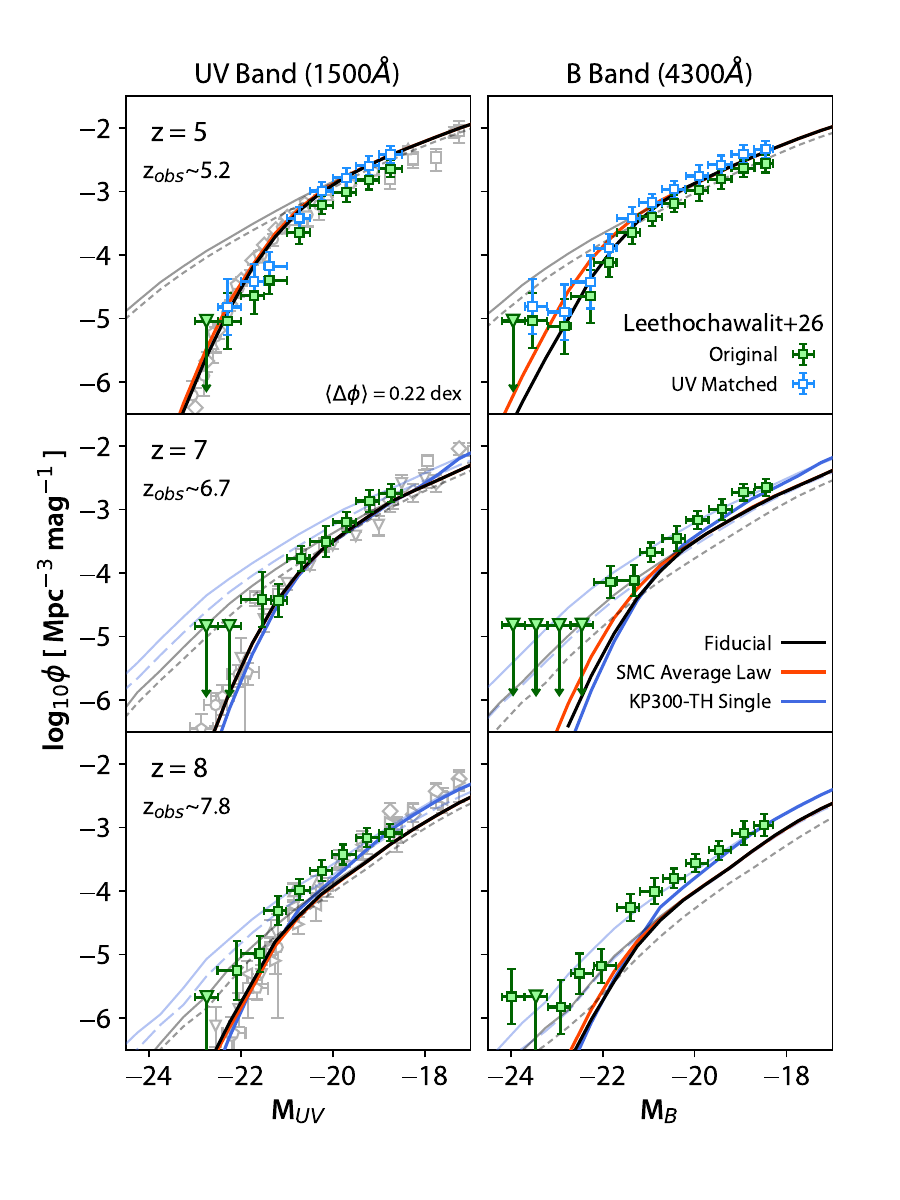}
    \caption{Comparison of the UV and B-band luminosity functions reported by \citet{Leethochawalit2026} (green data points) with our best fitted fiducial model discussed above. The color scheme used are identical to that of Figure~\ref{fig:ninja_lf_comp}. Blue data points shown in blue are measurements by \citet{Leethochawalit2026} scaled by 0.22 dex (see section \ref{sec:BLF} for details). This is to mainly illustrate that our model predicts the ratio of UV and B-band luminosity functions well at $z\sim$5. Short- and long-dashed curves are predicted luminosity functions considering only the stellar contribution to the continuum light.
    }
    \label{fig:ninja_lf_b_band}
\end{figure}

Independent constraints on the spectral energy distribution of galaxies and dust correction can be obtained by comparing luminosity functions in other wavelength ranges predicted by our simulations, which are calibrated using the observed UVLFs. It has been suggested that
LF measured in the rest frame optical wavelength (say in B-band)
is less susceptible to bursty star formation activities and dust attenuation and thereby will provide a less biased evaluation of galaxy luminosity functions and nature of dust at early cosmic times.

Recently \citet{Leethochawalit2026} have reported the rest frame B-band luminosity  function of galaxies over the redshift range $5\le z\le 9$. They found that the redshift evolution of the B-band luminosity function is stronger than that of the UVLF.
However, both UV and B-band luminosity functions are found to be evolving slowly with redshift compared to what has been predicted by simulations.  They also concluded that no single existing simulations produce all the trends found by them.
Their UVLF roughly follows the UVLF that we use to calibrate our dust models for $z\sim$ 7 and 8 (Figure~\ref{fig:ninja_lf_b_band}). However, their UVLF for $z = 5.2$ is found to be lower than the values from the literature at $z\sim5$ used in Figure~\ref{fig:ninja_LF}. \citet{Leethochawalit2026} attributed this to a stricter photometric redshift selection window adopted (see section 4.1 in their paper) for details.  

In this section , we compare our predicted UVLF and B-band luminosity functions for the best fit values of $\epsilon$ with those reported by \citet{Leethochawalit2026}.  As discussed before, dust effects are important only at high luminosities. Therefore, the predicted faint end B-band luminosity function,obtained by fitting the UV luminosity function, depends mainly on the {intrinsic} spectral energy distribution, which in the present case depends mainly on the assumed IMF and the relative contribution of diffuse emission at different wavelengths. 

As can be seen in Figure~\ref{fig:ninja_lf_b_band},
our fiducial model (black curves) consistently over-predicts both UVLF and B-band luminosity compared to the observations (green points with error-bars) for $z\sim$5 (top panel in Figure~\ref{fig:ninja_lf_b_band}). As discussed above, the UVLF values of \citet{Leethochawalit2026} are systematically lower than the rest of the measurement by 0.22 dex in the faint end ($M_\text{UV}>-20.5$). 
If we apply this off-set to their measurements at all magnitudes for both UV band and B band (shown in blue points) we find our model predicted B-band luminosity function nicely follows observed points corrected for this off-set. This confirms that our fiducial model consistently reproduces the observed ratio of UVLF to the B-band luminosity function at a given luminosity.   It is also evident from this figure that this ratio can not be reconciled if we consider only the stellar contribution to the continuum light (gray thin dashed curves).

\begin{figure*}
    \centering  
        \centering
        \includegraphics[trim={0cm 0.8cm 0cm 0cm}, clip, width=\linewidth]{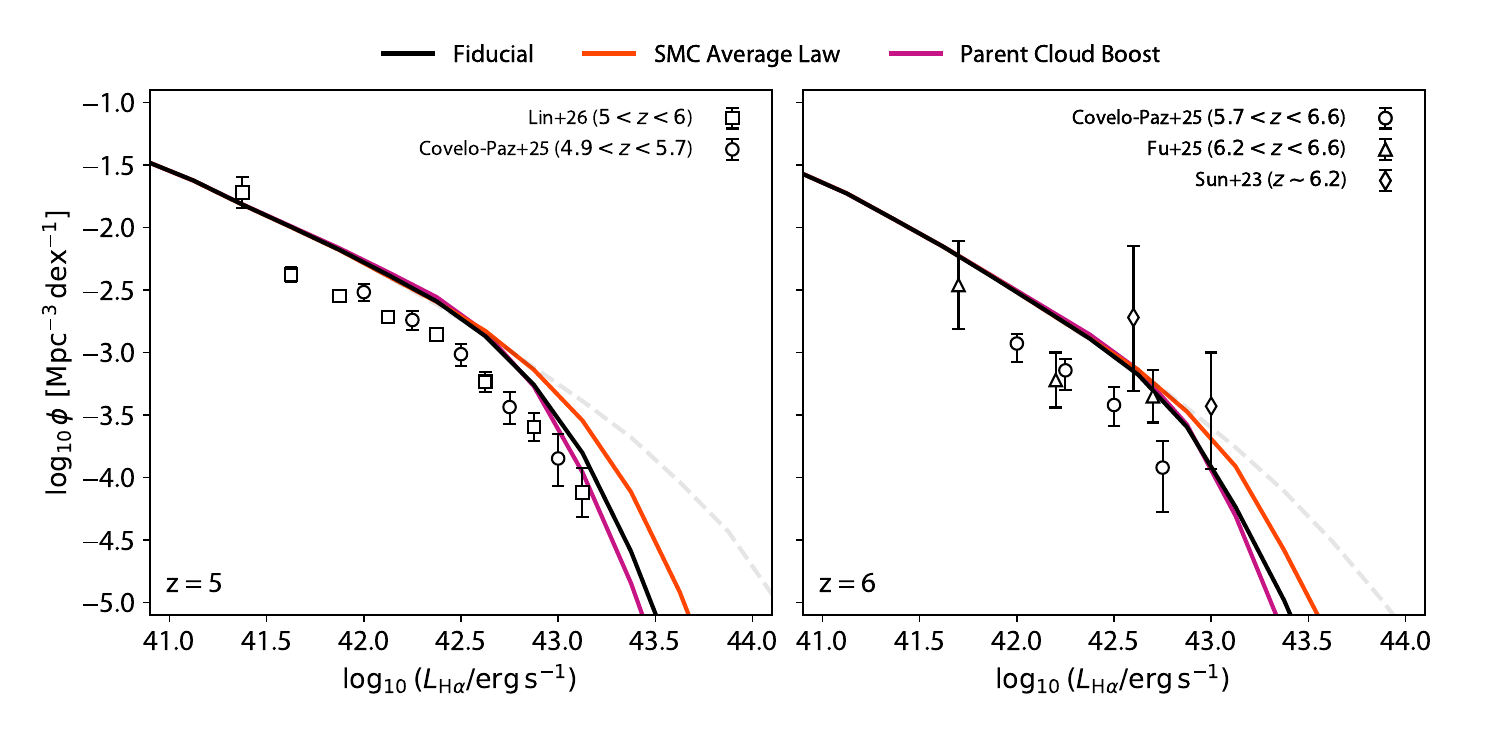}
    \caption{Comparison of observed H$\alpha$ luminosity function with the predictions from our NINJA simulations fo fiducial model (black), model that uses SMC extinction law (red) and with birth cloud attenuation(pink) for $z\sim5$ (left panel) and $z\sim$6 (right panel).  The data points for $z\sim5$ are from \citet{Lin2026} and \citet{Covelo-paz2025}. In the case of $z\sim6$ we use data points from \citet{Sun2023, Fu2025} and \citet{Covelo-paz2025}.
    The gray dashed curves in  both the panels are  the dust uncorrected H$\alpha$ luminosity functions from our simulations.} 
            \label{fig:ninja_HaLF}
\end{figure*}

The {orange-red} curves in this figure show our predictions for  the SMC extinction curve. It is clear that even though the UVLF for both attenuation curves are similar, we do see the differences in the high-luminosity end in the case of B-band. While the shifted data (i.e blue points) are well reproduced by our fiducial model, the models using SMC extinction curve slightly over-produce the B-band luminosity function.  However, with the present data it will be very difficult to favor one over the other due to large observational errors.  Therefore, before concluding how well our models reproduce the observations, it will be important (i) to understand the reason behind the off-set seen in the UVLF between that of \citet{Leethochawalit2026} and rest, and (ii) also to improve the B-band luminosity function at the high luminosity end.

In the case of $z\sim7$ (middle panels) the UVLF reported by \citet{Leethochawalit2026} roughly follows the data used by us from the literature. As can be seen in the figure, our predicted B-band luminosity function is slightly below the observed one.  We also show the results when we use the KP300-TH IMF (blue curves).
This produces slightly higher B-band luminosity function, that is closer to the observations, compared to our fiducial model.
We can clearly see that the model predictions will be more inconsistent if we consider only the stellar continuum without adding the nebular contribution (sort and long dashed curves for fiducial and KP300-TH IMFs, respectively). However, only upper limits are available for the B-band luminosity function at higher luminosities (i.e ${M}_{\rm UV}<-22$ mag). Therefore, we are not in a position to distinguish between our fiducial model and the model that uses the SMC extinction curve.

At $z\sim8$, the observed UVLF by \citet{Leethochawalit2026} roughly follows other data from the literature. As discussed before in the low-luminosity end our simulations (i.e fiducial case) slightly under predict the UVLF compared to the observations. Therefore, we also see the predicted B-band luminosity function to have a similar off-set. This figure also shows our predicted UVLF and B-band luminosity when we use the KP300-TH IMF. They provide better fits to both luminosity functions than the Chabrier IMF. It is interesting to note even in this case the predicted B-band LF considering only the stellar contribution under-predicts the observations.

In summary, for NINJA simulations to simultaneously fit both UVLF and B-band LF  it is important to include the nebular contribution to the continuum. More accurately measured B-band luminosity function at the high luminosity end (through large volume surveys) can provide an interesting constraint on the nature of average extinction curve and its redshift evolution. Moreover, well measured LFs at the low-luminosity end can be used to provide constraints on parameters of the simulations and/or IMF.

\subsubsection{The H$\alpha$ luminosity function}
\label{sec:lf_halpha}

As we discussed before, the diffuse emission computed using \software{CLOUDY} for each star particle also contains nebular emission lines. We obtain the H$\alpha$ luminosity of a halo by combining contributions from individual star particles after applying dust attenuation using the best-fitting $\epsilon$ (given in Table~\ref{tab:lf_fit_results} required to fit the UVLF for our fiducial dust model.
In Figure~\ref{fig:ninja_HaLF}, we compare the H$\alpha$ luminosity functions available for two redshifts with our simulation results.

For $z\sim6$ we use the observed luminosity functions from \citet[][]{Sun2023,Covelo-paz2025} and \citet{Fu2025}. Among them, the data from \citet{Covelo-paz2025} is based on a large area survey (GOODS-N and GOODS-S fields) and provides a good representation at the high luminosity end. \citet{Fu2025} data are towards Frontier fields where lensing by the four foreground clusters allows one to probe the low luminosity end very well. Both data sets are consistent with each other. Our model predictions are systematically higher. Also, over the observed luminosity range the differences in dust implementations do  not significantly affect our predicted luminosity functions.

For $z\sim5$ we used the observed H$\alpha$ luminosity function from \citet{Covelo-paz2025} and \citet{Lin2026}. Both use the same data, while \citet{Covelo-paz2025} uses available JWST grism observations towards GOODS-N and GOODS-S, \citet{Lin2026} uses only GOODS-N data. They also use slightly different $z$ ranges (as indicated in the figure). The H$\alpha$ luminosity functions for both samples are consistent while the number of objects in their catalog differs due to different procedures adopted to generate the sample \citep[we refer the readers to section 3.1 of][for detailed comparison between the two samples]{Lin2026}. At the low-luminosity end, our predicted H$\alpha$ luminosity (unaffected by dust reddening) is higher than the observed value. Typically, our models predict 0.4 dex more $\phi$ compared to observations at a given H$\alpha$ luminosity. Alternatively, we need to dim the H$\alpha$ luminosity by 0.3 mag to be consistent with the observed luminosity function over the full observed range.

It is also evident that at the high luminosity end, model using the SMC extinction curve over-produces and two-component dust model under-produces the H$\alpha$ luminosity function compared to our fiducial model. However, the reddening applied is not sufficient to make our predictions consistent with the observed H$\alpha$ luminosity function. The completeness correction is significant for log~L$_{\rm H\alpha}\le41.75$ and \mab$\le-20$ mag.
We also notice that the UVLF obtained for the spectroscopic sample of \citet[][see their figure 7]{Covelo-paz2025} is consistent with our best fit UVLF. Thus, the mismatch between the observations and the simulations has to come from the modification to the H$\alpha$ luminosity.

Two modifications to our \software{CLOUDY} modeling can be useful in reducing the H$\alpha$ luminosity function. The first is to introduce dust in the H~{\sc ii} region. Note that we consider a dust free gas in our sub-grid model and the H$\alpha$ luminosity is sensitive to the assumed dust content \citep[See figure A8 of][]{Wilkins2020}. On the other hand, models used in the literature usually assume Orion type dust grain composition scaled by metallicity \citep[for e.g.][]{Wilkins2020,Vijayan2021}. The second possibility is to use a covering factor less than 1. Note that cosmological reionization models typically require $\sim$ 14-20\% of the H~{\sc i} ionizing photons to escape the galaxy at $z>6$ to be able to reionize the universe and explain the H~{\sc i} photoionizaiton rate at $3\le z\le 5$ \citep[see for example,][]{Khaire2016}.
Interestingly, to understand the observed trend in the equivalent width distribution of [O~{\sc iii}]+H$\beta$ (or H$\alpha$) on \mab, \citet{Endsley2024} have considered the \mab-dependent escape fraction of the UV ionizing photon at $z>6$. They needed escape fractions (as discussed in section 5.3 of their paper) typically in the range 0.07-0.35 (or covering factor in the range 0.65 to 0.93). Such values will make our H$\alpha$ Luminosity function consistent with observations.

Thus, it appears that inclusion of dust and the covering factor can help us produce a consistent H$\alpha$ luminosity function.  However, introducing them into the cloudy modeling will have effects on the predicted  strength of the continuum emission and other nebular emission lines. Therefore, we wish to address this issue very carefully in our upcoming paper, which will mainly focus on the effect of varying various simulation parameters.

Another possibility is to have $A_V^\text{BC}$ much higher than 2$\left<A_V^\text{ISM}\right>$ assumed here \citep[also see][]{Vogelsberger2020}. It should be noted that in {our simulations of} high redshift galaxies, the stellar metallicities are found to be higher than the average metallicity of the ISM (Appendix~\ref{sec:ninja-scaling-relations}). Therefore, the metallcity and dust-to-metal ratio in  the birth cloud can be much higher. Hence, it is reasonable not to relate the dust attenuation caused by the birth cloud and that of the ISM, as done in the case of \FLARES\ simulations \citep[See][]{Vijayan2021}.  Any modifications to $A_V^\text{BC}$ will have implications for \buv\ and the relationship between observed ${\rm E(B-V)_{star}}$ and ${\rm E(B-V)_{neb}}$ obtained using the Balmer ratio. We come back to these points when we discuss \buv\ (section~\ref{sec:colormag}) and the Balmer ratio (section~\ref{sec:Bal_ratio}).

In summary, the simulations and light generation models presented here over-produce the observed H$\alpha$ luminosity function at $z\sim5$ and 6. However, we believe that there is enough room in the model parameters that will enable us to address this issue. 
{We will analyse such scenarios in a future paper.}

\subsubsection{UVLF for
$z>10$ galaxies}

\begin{figure}
    \centering
    \includegraphics[bb=10 10 390 410, width=\columnwidth, clip=true]{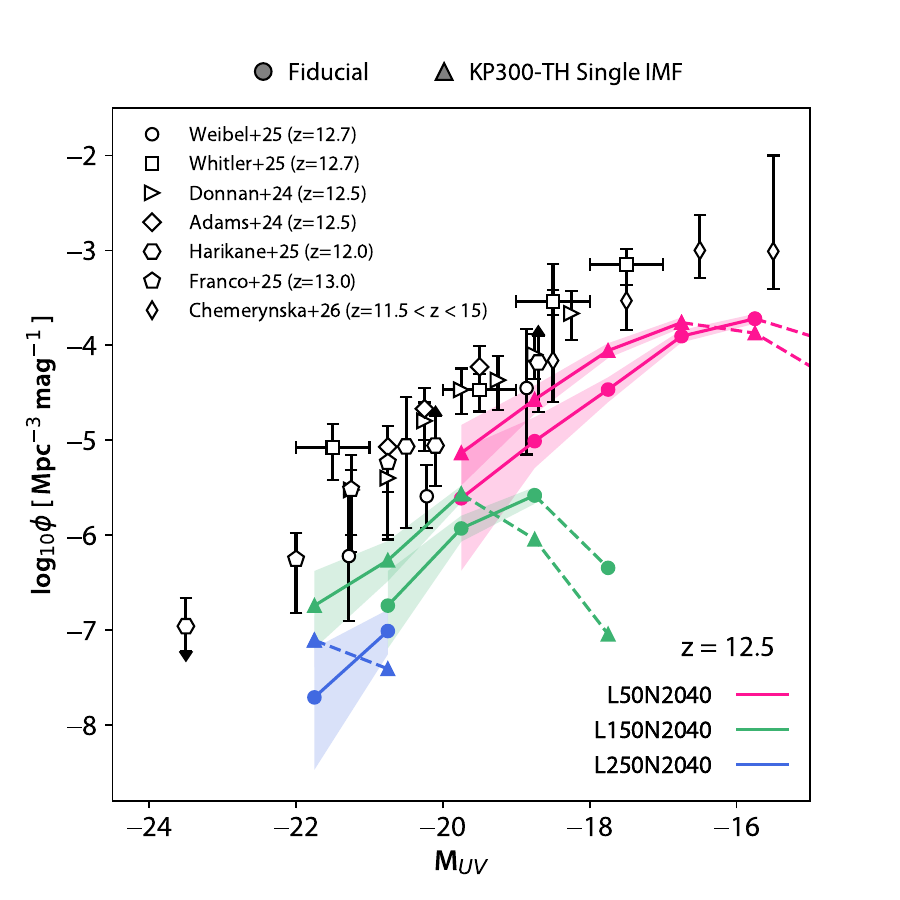}
    \caption{ Comparison of our UVLF at $z\sim 12.5$ predicted by the NINJA simulations with available observations from \citet{Weibel2025, Whitler2025, Donnan2024,Adams2024,Harikane2025, Franco2025} and \citet{Chemerynska2026}. Simulation results using Chabrier and KP300-TH are presented with filled circles and triangles respectively. No dust correction and incompleteness correction is applied. The incomplete range is indicated with dashed line.  }
    \label{fig:ninja_lf_z12p5}
\end{figure}

In this section, we mainly focus on the UVLF of galaxies in our simulations at $z\ge 10$.  In Table~\ref{tab:highz_gal_count} we summarize the number of galaxies (with at least 16 star particles) at $z\ge 10$ in the simulations  L50N2040, L150N2040, and L250N2040. The number of galaxies in our simulations dramatically falls beyond $z=10$ in particular both L150N2040 and L250N2040 have only one galaxy at $z\sim15$. This could be related to the completeness and resolution issues in the simulations that need to be addressed before drawing any conclusions on possible issues associated with LCDM simulations. 
So, here we focus mainly on UVLF at $z\sim12.5$ from our simulations.

The UV Luminosity functions obtained for these individual boxes without applying corrections for incompleteness and dust attenuation are shown in Figure~\ref{fig:ninja_lf_z12p5}. We present the results for two IMFs (Chabrier in filled circles and KP300-TH in filled triangles) without applying dust corrections. It is clear from the figure that the UVLF predicted by our fiducial model is systematically lower than the observed data points. This is consistent with what we have seen at $z\sim10$ (or at the low luminosity end of $z\ge7$). 
The UVLF evolution predicted by our fiducial model is more rapid than what is seen in the observations.

\begin{table}
    \centering
    \renewcommand{\arraystretch}{1.}
    \begin{tabular}{cccc}
         \hline
         $z$ & L50N2040 & L150N2040 & L250N2040 \\
         \hline
         10.0 & 2177 & 1795 & 615\\
         12.5 & 210 & 55 & 9\\
         15.0 & 15 & 1 & 1\\
         \hline
    \end{tabular}
    \caption{Number of halos with minimum 16 star particles for the three boxes at different redshifts.}
    \label{tab:highz_gal_count}
\end{table}

 On the other hand, when we use the KP300-TH IMF the situation is slightly better. For example, our simulated UVLF
 is consistent with the measurements reported in \citet{Weibel2025} (with a survey area of 0.28 deg$^2$) and \citet{Franco2025} (with a survey area of 0.54 deg$^2$). However, these predicted UVLF are systematically lower than the UVLF reported by \citet{Whitler2025} (with a net survey area of $\sim$ 0.05 deg$^2$) and \citet{Donnan2024} (with a net survey area of  0.1 deg$^{2}$). In the low luminosity end (i.e $-19\le$\mab$\le-17$) our model using KP300-TH IMF is consistent with the UVLF reported by \citet{Chemerynska2026} and UVLF limit based on spectroscopic data reported by \citet{Harikane2025}. However, less than that reported by \citet{Donnan2024} and \citet{Whitler2025}. 
 
 In summary, to closely match the various observations shown in Figure~\ref{fig:ninja_lf_z12p5}(or the median of available observations), we need to explore the varying different parameters in our simulations, which will be presented in our upcoming paper. 
However, it is also evident that the UVLF observations available at $z\sim12.5$ have much larger scatter at a given \mab. 
 Since most of these measurements are based on photometric redshifts, various steps like candidate selection, signal to noise cutoff, interloper identification, and understanding of incompleteness using simulations can lead to scatter in the observed UVLF. 
This has been acknowledged by various authors. In addition, cosmic variance could be at play and the data will converge once UVLF is obtained over large volumes \citep[see for example,][]{McLeod2026}.

\subsection{Dust in high-\texorpdfstring{$z$}{z} galaxies}

\label{sec:dust_mass}

\begin{figure}
    \centering
    \includegraphics[bb = 10 10 395 500,clip=true,width=\linewidth]{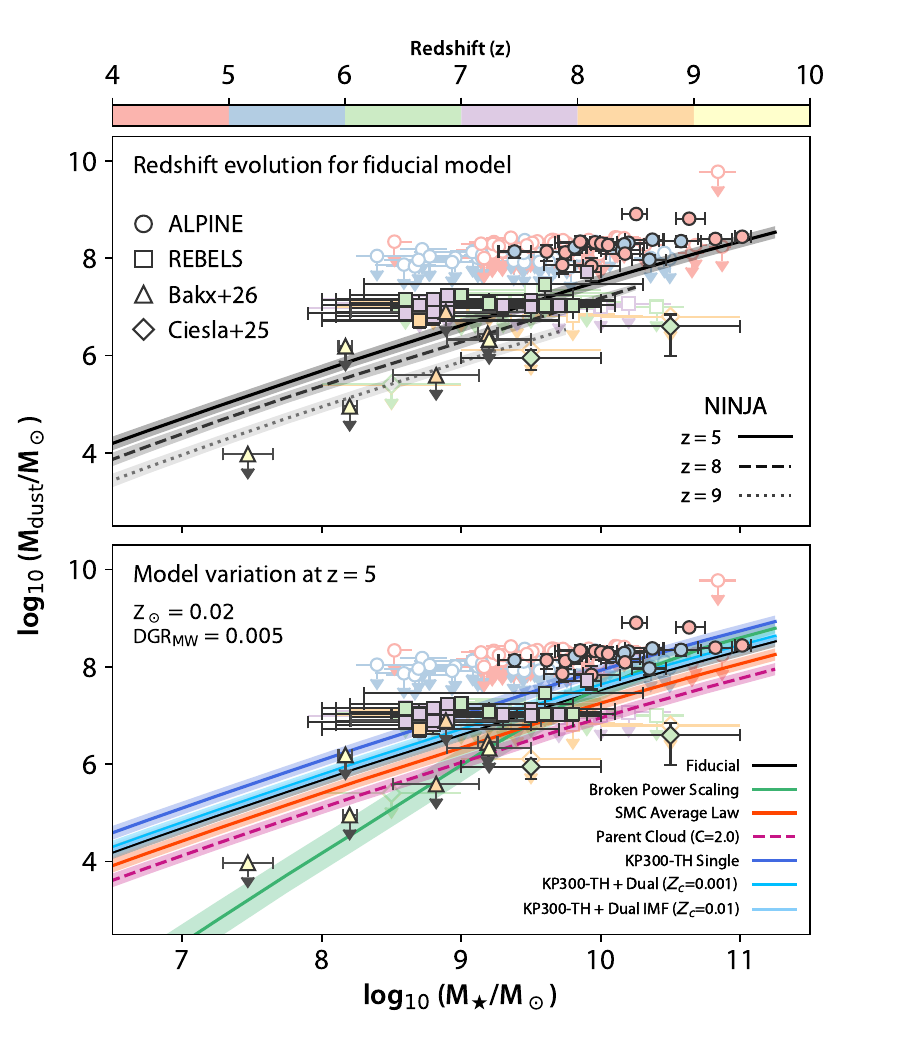} 
    \caption{Mean dust mass (${\rm M_{dust}}$) – stellar mass (${\rm M_\star}$) relation for simulated galaxies under different light synthesis models calibrated by fitting the UVLF (see section ~\ref{sec:UVLF} for details). The lower panel illustrates the impact of model variations at $z=5$ (with the colour scheme similar to that used in Figure~\ref{fig:ninja_lf_comp}), while the upper panel shows the redshift evolution for the fiducial model. The shaded region indicates the $1\sigma$ scatter. Dust masses are computed assuming a local dust-to-gas ratio of $1/200=0.005$ at the solar metallicity of $Z_\odot=0.02$, and incorporating model-dependent metallicity scaling (as described in section~\ref{sec:dust}). Observational data, color-coded by redshift, are derived from the ALPINE catalog \protect\citep{Bethermin2020} and the REBELS catalog \citep{Inami2022}, with dust masses estimated using a modified blackbody as described in \citep{Pozzi2021}. We also plot the measurements presented by \citet{Bakx2026} for $8\le z\le 10$ and stacking measurements from \citet{Ciesla2025}.
    }
    \label{fig:stellar-to-dust-mass-relation}
\end{figure}

The relationship between dust mass (${\rm M_{dust}}$) and ${\rm M_\star}$ and its redshift evolution provide important constraints on various dust production mechanisms and their redshift evolution \citep[see for example,][]{Popping2017}. 
In Section~\ref{sec:lf_dust_var}, we explored a range of stellar population synthesis models and found that all reproduce the UV luminosity function (UVLF), with deviations from the fiducial model remaining within $\sim 10\%$ (see Figure~\ref{fig:ninja_lf_comp}). However, each model requires a different dust-to-hydrogen ratio, $\epsilon$, in order to match the observed UVLF (see Table~\ref{tab:lf_fit_results}). For each galaxy in the simulation, we compute the corresponding dust mass for a given synthesis model by assuming a local dust-to-gas ratio (a typical Milky Way value of $1/200 \approx 0.005$ at solar metallicity) and applying the model-specific metallicity scaling as described in Section~\ref{sec:dust}. In Figure~\ref{fig:stellar-to-dust-mass-relation}, we compare the resulting ${\rm M_\star}$ - ${\rm M_{dust}}$ relation with observational data.

For high-$z$ observations we mainly use ALMA measurements for ALMA ALPINE \citep[ALMA observations of 118 spectroscopically confirmed star-forming galaxies at $4.4\le z\le 5.9$ as reported by][with 23 firm detections in the mm continuum]{Bethermin2020} and ALMA REBELS \citep[ALMA observations of 49  \mab$ < -21.3$ mag galaxies at $z > 6.5$ as reported by ][with 18 firm detections in the mm continuum]{Inami2022}. We convert continuum measurements and upper limits in the millimeter range into ${\rm M_{dust}}$ and upper limits in ${\rm M_{dust}}$ using the relationship \citep[][]{Pozzi2021},
\begin{align}
  {\rm   M_\text{dust} = \frac{D_L^2 S_{\nu_{\text{obs}}}}{(1+z)\kappa_\nu B_\nu(T_{dust})}}
\end{align}
where $\nu$ and $\nu_{\rm obs}$ are the observed and rest-frame frequencies (with $\nu_{\text{rest}}=\nu_{\text{obs}}(1+z)$), B$_{\nu}$(T) is the Planck function, ${\rm D_L}$ the luminosity
distance, $\kappa_\nu = \kappa_0 (\nu/\nu_0)^\beta$ cm$^{-2}$ ${\rm  gm^{-1}}$ is the grain absorption cross section per unit mass, and
${\rm S_{\nu_{obs}}}$ is the observed flux corresponding to a rest-frame frequency
at which dust can be considered optically thin.
The derived ${\rm M_{dust}}$ depends very much on our choice of dust emissivity index $\beta$ and dust temperature ${\rm T_{dust}}$.
Following \citet{Inami2022} we use  $T_\text{dust}=45\rm{K},~\beta=2, ~\kappa_0 = 10 \text{ cm}^2\text{g}^{-1}, \nu_0 = 1.9~\rm{THz} $ for the ALMA REBELS data. In the case of ALMA ALPINE data 
we have used  ${\rm T_\text{dust}=25\rm{K}, \beta=1.8, \kappa_0 = 4 \text{ cm}^2\text{g}^{-1}, \nu_0 = 1.2~\rm{THz} }$ as in \citet{Pozzi2021}. The stellar mass measurements for galaxies in the ALMA REBELS catalog are reported in \citet{Bowler2024} and the same for the ALMA ALPINE are from \citet{Faisst2020}.  

In addition to these two data sets, we have also included the upper limits reported by \citet{Bakx2026} for objects in the redshift range $8<z<10$, where $\beta =2 $ and ${\rm T_{dust}} $= 50 K are assumed. Recently, \citet{Ciesla2025} have reported stacked millimeter fluxes or upper limits for galaxies at $z>6$ in the JADES sample binned in ${\rm M_\star}$.  We converted these into upper limits on ${\rm M_{dust}}$ using $\beta =2 $, $\kappa_0=10.4$ cm$^{2}$ g$^{-1}$, $\nu_0 = 1.0$ THz and ${\rm T_{dust} }$= 50 K for the redshift bins (i.e $ 6 \le z \le 7$ and $7\le z \le 9$). For the stellar mass range, 8$\le {\rm log~M_\star}\le 9$ is the upper limit derived on ${\rm M_{dust}}$ is $\simeq 3\times 10^5$ M$_\odot$ for both redshift ranges.

In the bottom panel of Figure~\ref{fig:stellar-to-dust-mass-relation}, we compare the predicted relationship between ${\rm M_{dust}}$ and ${\rm M_\star}$ by our models (discussed in sections~\ref{sec:lf_dust_var} and ~\ref{sec:lf_imf_var}) for $z=5$. Galaxies with clear ${\rm M_{dust}}$ measurements have high ${\rm M_\star }$ and galaxies with log~${\rm M_\star}\le 9.5$ only have upper limits for ${\rm M_{dust}}$.   Although this is consistent with the existence of correlation between ${\rm M_{dust}}$ and ${\rm M_\star}$ we are unable to constrain the slope (which, in our model, depends on how dust-to-metal ratio is related to metallicity) from these individual measurements and upper limits.

For a given stellar mass, the mean values of ${\rm M_{dust}}$ predicted by different models vary a lot. The highest dust masses are obtained for the model that uses the “KP300-TH” IMF, compared to our fiducial model. In contrast, adopting the “SMC” attenuation curve results in lower predicted dust masses at fixed ${\rm M_\star}$.  The ${\rm M_{dust}}$ shown for the two-component dust model (megenta color dashed curves) should be considered as a lower limit, as we could not compute the dust mass contributed by the birth clouds self-consistently.
Interestingly, all models that assume the dust-to-metal ratio to scale with metallicity yielded a similar slope for the ${\rm M_\star - M_{dust
}}$ relation. The differences between these models are mainly reflected in normalization, set by the best-fit values $\epsilon$ (see Table~\ref{tab:lf_fit_results}). 
On the other hand, in the model where the dependence of dust-to-metal ratio on metallicity is a double power-law, we see a steeper slope for the relationship between ${\rm M_\star}$ and ${\rm M_{dust}}$.

As can be seen, all ALMA ALPINE measurements ${\rm M_{dust}}$ are higher than the mean values predicted by our fiducial model. 
For the simulation parameter considered here only the model using the "KP300-TH” IMF produces ${\rm M_{dust}}$ closer to what has been observed at the high ${\rm M_\star}$ end. However, the predicted values of ${\rm M_{dust}}$ for "KP300-TH" are higher than  inferred ${\rm M_ {dust}}$ using the stacking results of \citet{Ciesla2025}.
Thus, it appears that to explain the ALMA ALPINE observations at high ${\rm {M_\star}-end}$, we would require dust in excess of what is required to fit the  UVLF in our fiducial models. In addition, one may favor a flatter attenuation curve compared to that of the SMC. Here, we assume that the ALMA ALPINE measurements represent that of the general population of galaxies.
 On the other hand, to explain the observational upper limit on ${\rm M_{dust}}$ at low-${\rm M_\star}$, the broken power law scaling between the dust-to-metal ratio and metallicity may be favored.

In the top panel of Figure~\ref{fig:stellar-to-dust-mass-relation} we show the redshift evolution of the ${\rm M_{dust}}$ - ${\rm M_\star}$ relationship for our fiducial model.  As suggested by the redshift evolution of $\epsilon$ (see Table~\ref{tab:lf_fit_results}),
for a given ${\rm M_\star}$ the predicted dust mass decreases rapidly with increasing redshift. The plotted observations also indicate the reduction in ${\rm M_{dust}}$ with increasing $z$.  
It is evident that the ALMA REBELS measurements (for $z>6.5$) are also higher than the predictions of our model. Our curves match only a few measurements at the high ${\rm M_\star}$ end. 
On the other hand, we do see that the upper limits for the $z>8$ galaxies from \citet{Bakx2026} and \citet{Ciesla2025} are consistently lower than our model predictions. Thus, it appears that the observed dust evolution may be much stronger than predicted based on the evolution of $\epsilon$. 
As in simulations the amount of dust required depends on the efficiency with which one forms stars, the inferred redshift evolution will depend on the parameters used in our simulations. Therefore, we will come back to this issue in our upcoming paper where we study simulations that are run with different sets of parameters. 

\begin{figure*}
    \centering
    \includegraphics[bb = 18 17 820 485, clip=true,width=\textwidth]{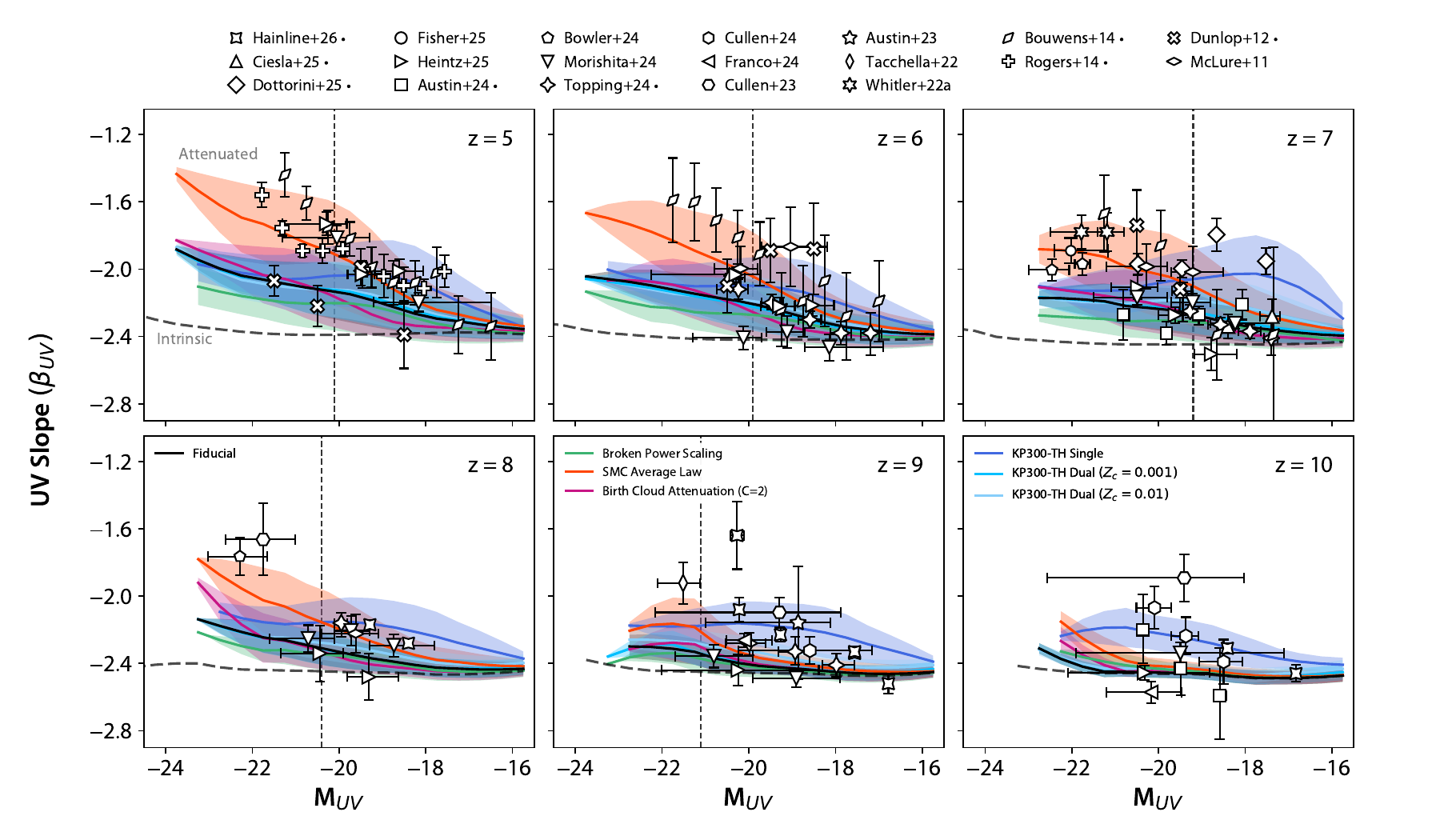}
    \caption{The UV-slope \buv\ vs. \mab\ for different redshift bins.  The observational points are from \citet{Hainline2026, Ciesla2025, Dottorini2025,Fisher2025, Heintz2025,Austin2024,Bowler2024,Morishita2024,Topping2024,Cullen2024,Cullen2023,Franco2024, Austin2023, Tacchella2022,Whitler2022a, Bouwens2014, Rogers2014, Dunlop2012,McLure2011}. We only use the mean value of \buv\ in different \mab\ bins. When the binned measurements are already available (author name in legend with a '$\bullet$') we directly use them. Otherwise we generate these binned measurements using standard methods.
    In all plots the long dashed curve provides the expected \buv\ for the intrinsic continuum for our fiducial model. The color scheme used for different models are same as in Figure~\ref{fig:ninja_lf_comp}. The vertical dashed line is same as in Figure~\ref{fig:ninja_LF}.
    }
    \label{fig:ninja_colmag}
\end{figure*}

On the observational side, it is important to note that millimeter continuum emission is detected in only 19\% of sources in the ALMA ALPINE sample and 37\% in ALMA REBELS. Moreover, these samples are preselected for ALMA observations, which may have introduced an additional bias. As a result, the mean values derived from these measurements are unlikely to be representative of the overall galaxy population, unlike the averages predicted by our models.
This is clearly indicated by the upper limits on ${\rm M_{dust}}$ obtained for different ${\rm M_\star}$ bins using stacking measurements of \citet{Ciesla2025}.
We also find that the fraction of sources with continuum detections increases with stellar mass (${\rm M_\star}$), being higher at larger ${\rm M_\star}$. This trend is consistent with the expected correlation between stellar mass and dust mass (${\rm M_{dust}}$). However, the current individual upper limits on ${\rm M_{dust}}$ are not sufficiently stringent to discriminate between the two dust-to-gas (DTG)–metallicity scaling relations adopted in our models. However, the inferred upper limits based on stacking may favor two power law model.
Therefore, systematic measurements of ${\rm M_{dust}}$ over a wide range of ${\rm M_\star}$
at a given $z$ are needed to put stringent constraints on the parameters of the dust models used here.

\subsection{\texorpdfstring{\mab}{MAB} vs \texorpdfstring{$\beta_{\rm UV}$}{betaUV} relationship} 
\label{sec:colormag}

The UV continuum slope (\buv) of a galaxy is shaped by the stellar age, metallicity, and continuum attenuation (i.e dust modeling and attenuation law used) for a given IMF. Discussions presented in section~\ref{sec:UVLF}, show that the best fit to UVLF at the high luminosity end is obtained using dust reddening that increases with metallicity and therefore on stellar mass and luminosity. It is also expected that the spectral slope can distinguish between different attenuation curves used in our simulations (as shown in section~\ref{sec:BLF}).
 Therefore, we expect a correlation between the UV-slope (\buv) and \mab\ and the relationship should evolve with redshift and should depend on the assumed attenuation curve.  By comparing the observed relationship between \buv\ vs \mab\ (or stellar mass) with the model predictions, we should be able to distinguish between various dust models and IMFs discussed above.

The relationship between \buv\ and \mab\  predicted by different dust extinction models that fit the UV-luminosity function well 
are shown in Figure~\ref{fig:ninja_colmag}. 
It is evident from the figure that at the faint \mab\ end (i.e fainter than the verical lines shown in Figure ~\ref{fig:ninja_LF}) \buv\ predicted by our models are nearly flat and mainly sensitive to the assumed IMF.
In the fainter end, the \buv\ values are systematically higher (i.e redder continuum with $\delta$\buv $\sim$ +0.2) when we use the KP300-TH IMF. 
It is also evident from Figure~\ref{fig:ninja_colmag} that the slope (i.e $d$\buv/$d$\mab) of the predicted curves, over the full \mab\ range considered here, are not very sensitive (i.e fall within $\pm0.01$) to inclusion of birth cloud attenuation or metallicity scaling used to obtain the dust abundance in our post processing step. We notice that the SMC extinction law produces a steeper slope for the \buv-\mab\ plot compared to what we get for the Calzetti extinction law. 

Various available observation points are also shown in Figure~\ref{fig:ninja_colmag}. The points and errorbars
are measurements binned over \mab. The legends in the figure with "$\bullet$" are the references, where the binned measurements are provided. In the remaining cases, we have obtained the binned measurements using standard procedures (details can be seen in Appendix~\ref{sec:obs_data}). 
We note that individual measurements have a very large scatter at any given \mab.
Observational measurements of \buv \ are susceptible to various systematics \citep[see][for details]{Bouwens2014}. Although the measurements of $\beta$ at high-$z$ using JWST photometry and spectroscopy are rapidly increasing, when we use them to generate population statistics, we must exercise caution bearing in mind various issues discussed in \citet{Bouwens2014}.

First, we focus on $z$ = 5 and 6 where the measurements are primarily based on HST. Here, we base our discussion mainly on the data presented by \citet{Bouwens2014}, while also showing all the other available observations in Figure~\ref{fig:ninja_colmag}. \citet{Bouwens2014} have shown a correlation between \mab\ and \buv\ that becomes weaker at the low luminosity range (faintward of -19 mag) compared to the high luminosity range. Their data seem to prefer a piecewise-linear relationship compared to a single linear relationship. Most of the other observational data that we have added also follow this trend (see Figure~\ref{fig:ninja_colmag}).
It is clear from the figure~\ref{fig:ninja_colmag}, our model predicted \buv\ values are consistent with their measurements at \mab$\ge-18$ mag.

\begin{figure*}
    \centering  
    \begin{subfigure}[t]{0.48\textwidth}
        \centering
        \includegraphics[bb = 10 15 420 410,width=\linewidth,clip=true]{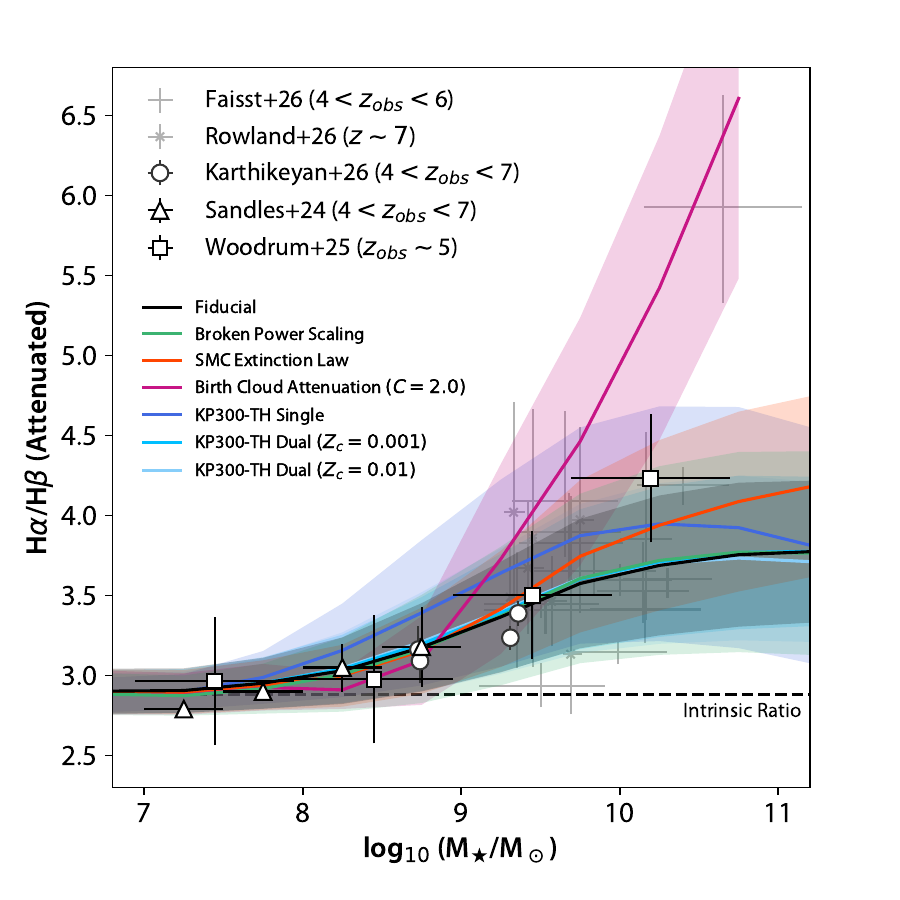}
        \label{fig:ninja_ms_vs_hahb}
    \end{subfigure}
    \hfill
    \begin{subfigure}[t]{0.48\textwidth}
        \centering
        \includegraphics[bb=15 15 425 410,clip=true,width=\linewidth]{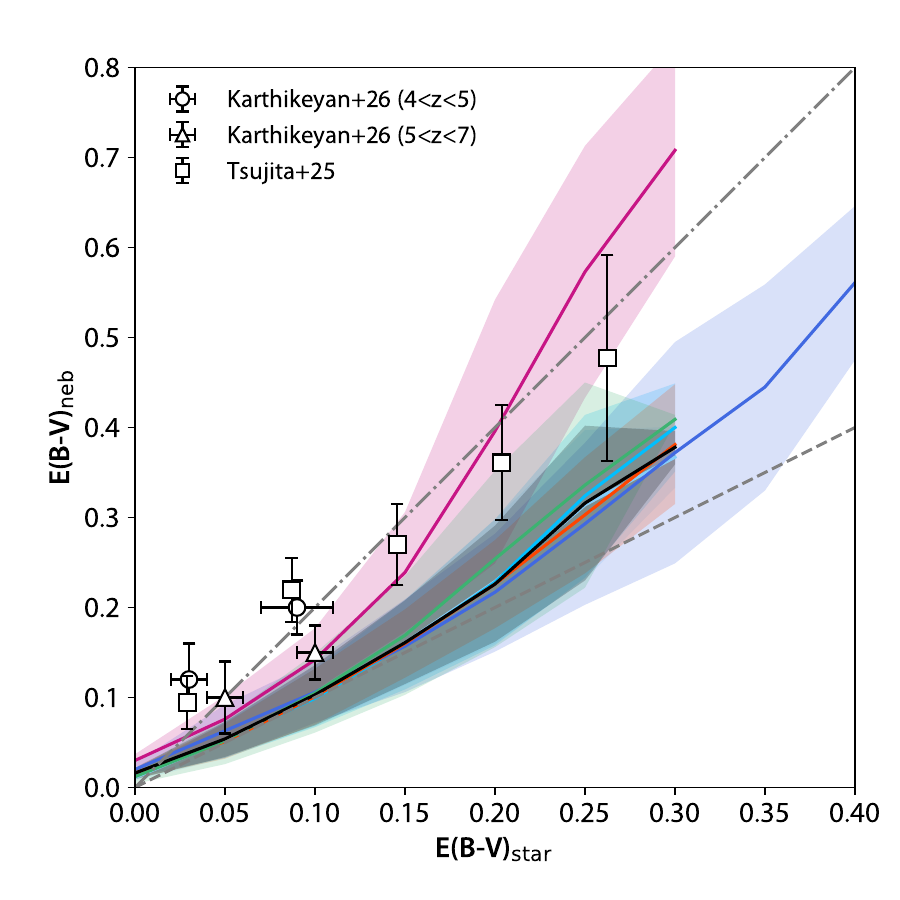}
        \label{fig:ninja_ebv_star_vs_neb}
    \end{subfigure}
    
    \caption{{\it Left Panel:} Balmer ratio (i.e H$\alpha$/H$\beta$) as a function of stellar mass. Our model predictions for different dust implementation considered in this work for $z=5$ are plotted on top of available measurements for galaxies at $4\le\ z\le7 $. We use the same color scheme as in previous figures.  Our model predictions have a larger deviations at log (${\rm M}_\star/{\rm M}_\odot$) $\ge$ 10, where at present the available observations are sparse. In low stellar mass end all our dust implementations are consistent with the available average measurements. {\it Right panel:} The ${\rm E(B-V)_{neb}}$ obtained from the Balmer decrement is plotted against ${\rm E(B-V)_{star}}$ obtained from the continuum reddening. The dashed lines provide one-to-one relationship and the dot-dashed line shows 1:2 relationship. Most of the binned data points are consistent with the dot-dashed line. In our case only model with birth cloud attenuation produce higher values of ${\rm E(B-V)_{neb}}$ compared to ${\rm E(B-V)_{star}}$.
    }
    \label{fig:balmer_ratio}
\end{figure*}

In both redshift ranges, our models using the Calzetti attenuation curve tend to have a flatter \mab-\buv\ relationship
compared to the one that uses the SMC extinction curve.  
The deviation 
between different dust models that use the Calzetti attenuation curve with respect to our fiducial model is less than the difference between our fiducial model and the one that uses the SMC extinction  curve. Within the observational uncertainties, the observations of \citet{Bouwens2014} favor the model that uses the SMC extinction curve (or the attenuation curve slightly steeper than that of SMC). 
Thus, available measurements seem, on average, to favor a steeper extinction curve compared to the Calzetti attenuation law at $5\le z\le6$. Alternatively, if we wish to produce the observed relationship using the Calzetti attenuation law, then we need to have more ${\rm A_V}$ than what we obtained in the present case. This may mean producing more intrinsic UV light and dust compared to the models discussed here.
As discussed before (see section~\ref{sec:BLF}, \ref{sec:lf_halpha}) when we use the SMC extinction curve, our model predicts systematically higher values for the B-band, the H$\alpha$ luminosity function, and lower ${\rm M_{dust}}$ (see section~\ref{sec:dust_mass}) compared to the predictions of our fiducial model.

When we consider $z\sim7$, at \mab$\ge$-19 there is no significant correlation between \mab\ and \buv. Although most of the data are internally consistent with each other in this luminosity range, we do see that systematically higher values were measured by \citet{Dottorini2025}. Since the number of sources involved in this sample is much less, we give more weightage to data from other sources. It seems that all our models produce consistent \buv \  in this luminosity range, with the model using the KP300-HT IMF producing systematically redder slopes. Even in this redshift range \buv\ measured at the high luminosity end are better represented by models using SMC like extinction curve compared to our fiducial model.

The results for $z\ge7$ are summarized in the bottom panels of Figure~\ref{fig:ninja_colmag}.  For $z\sim8$ available observations at the low luminosity end (not affected by dust) are consistent with all our models within observational uncertainties. The curves differ significantly at the high-luminosity end. The two \buv\ measurements presented in the figure are from \citet{Cullen2023} and \citet{Bowler2024}. These are redder than what is predicted for the Calzetti attenuation curve. To explain this, one will need a steeper attenuation or extinction curve like the SMC one.
As can be seen from Table~\ref{tab:lf_fit_results}, the reddening effects are not that important for $z\ge9$. For $z$ = 9 and 10, we note that there is a large scatter in the observed points. On average, the measured values of \buv \ tend to be slightly higher than the predictions of the fiducial model, but consistent with our predictions for the KP300-TH IMF. Such a case will favor the redshift evolution in the IMF as suggested by the luminosity evolution discussed in section~\ref{sec:UVLF}. 

However, before drawing firm conclusions, it is important to consider various aspects. (i) It is well documented that measurements \buv \ are susceptible to systematic uncertainties \citep[see][for details]{Bouwens2014}. Most of the measurements reported on the basis of the JWST measurements do not include systematic uncertainties. (ii) Even though we use JWST measurements from various references, we have not cross-checked for duplications of objects between the different samples used. This study is needed to avoid biased estimation of the relationship between \buv\ and \mab . (iii) It is well documented that individual measurements at a given \mab\ have a large scatter compared to the uncertainties associated with the binned measurements. 
Independent constraints on the nature of dust attenuation curves can be obtained using the direct recovery of attenuation curves from individual galaxies and also from the observed IRX-$\beta$ relationship. 

\subsection{Balmer Ratio \& {Color Excess}}
\label{sec:Bal_ratio}

The deviation Balmer ratio (i.e ${\rm H\alpha/H\beta}$) from its standard ratio established from the recombination rate (i.e Balmer decrement) is a good indicator of dust attenuation.
It is well established for the local universe and at $z\sim2$ \citep[see][]{Shapley2022} that the Balmer ratio has a strong correlation with the stellar mass.  Recent measurements using JWST spectra have found that this relationship does not evolve very strongly with redshift \citep[][]{ shapley2023, Sandles2024, Woodrum2025, Faisst2026,Karthikeyan2026}. All of these measurements indicate a large scatter in the H$\alpha$/H$\beta$ ratio at a given ${\rm M_\star}$. However, the running median of these individual measurements and measurements based on stacked spectra binned in redshift and ${\rm M_\star}$ are consistent with the trend shown at low-$z$ data within measurement uncertainties.  

In the right panel of Figure~\ref{fig:balmer_ratio}, we plot our model predictions of this ratio together with the available observations (binned medians) for $5<z<7$. For $7\le {\rm log~ M_\star} \le 10$, our model predictions for $z\sim5$ are consistent with available observations. Even the model with the birth cloud attenuation shows large attenuation only at very large ${\rm M_\star}$ where observations are sparse as of now.  Also we wish to remind the readers that ${\rm M_\star}$ in our case is a simple integration of mass of all the stars in the halo while in the case of observations it is obtained through SED fitting which depends very much on the assumptions related to the star formation history.

In the right panel of Figure~\ref{fig:balmer_ratio} we show the relationship between E(B-V) obtained using the Balmer ratio (i.e ${\rm E(B-V)_{neb}}$) and that obtained using continuum reddening (i.e ${\rm E(B-V)_{star}}$). \citet{Calzetti2000} have defined $f = {\rm E(B-V)_{star}/E(B-V)_{neb}}$ and found it to be 0.44$\pm$0.03. This was the motivation behind introducing additional attenuation in the case of young starts as discussed in Section~\ref{sec:dust}.
\citet{Woodrum2025} have reported a large scatter in the relationship between ${\rm E(B-V)_{neb}}$ and ${\rm E(B-V)_{star}}$ for $3 \le z\le 7$ (see their figure~5). They also noticed that their measured ${\rm E(B-V)_{star}}$ values are lower and ${\rm E(B-V)_{neb}}$ values are much higher. This led them to suggest that the values of $f$ may be evolving with redshift.
Recently \citet{Tsujita2026} have reported high spatial resolution spectroscopic measurement of ${\rm E(B-V)}$ for nebular and stellar components for 18 $z\sim6$ galaxies. Although spatially average individual measurements (as well as pixel by pixel measurements) show a large scatter, on average they reported $f=0.51^{+0.04}_{-0.03}$. The binned average of the pixel-by-pixel measurements are shown in Figure~\ref{fig:balmer_ratio}. These are consistent with $f=0.5$ (the dot-dashed line in the figure). \citet{Karthikeyan2026} reported measurements for galaxies at $2.7\le z \le 7$. They also notice that while there is a large scatter in this relationship a clear trend is seen when stacking based measurements are used. We show their measurements based on stacking for two redshift bin as reported in their Table~1. It is evident that $f$ for the highest redshift bin lies between 1 and 0.5. Clearly more observations are needed to get a clear trend. 

It is clear from the figure that our models (other than the one with the birth cloud attenuation) are consistent with the one-to-one relationship for ${\rm E(B-V)_{star}}<0.2$ albeit with a scatter that systematically increases with increasing ${\rm E(B-V)_{star}}$. As expected, our model using the birth cloud attenuation produces systematically higher ${\rm E(B-V)_{neb}}$ for a given ${\rm E(B-V)_{star}}$ compared to the rest of the models.
However, even these models produce $f$ values between 0.5 and 1.0 for ${\rm E(B-V)<0.15}$. It is evident from Figure~\ref{fig:balmer_ratio} that the $f$ values much less than 0.5 (as suggested by the individual measurements reported in \citet{Woodrum2025})  are not easily produced in our models.  
In order to produce $f\le0.5$ in galaxies with low ${\rm E(B-V)_{star}}$ we may have to enhance the ${\rm A_V^{BC}}$ more than what we have assumed here (i.e $>2\times \langle A_V^{\mathrm{ISM}}\rangle $). As discussed before (see section~\ref{fig:ninja_HaLF}) such a modification will also reduce the difference between the observed and predicted H$\alpha$ luminosity function.

\section{Summery \& Discussions}
\label{sec:summary}
In this work, we have introduced the NINJA suite of cosmological hydrodynamical simulations aimed at studying the formation and evolution of galaxies in the first billion years from the Big Bang, as probed by ever growing observations using JWST and ALMA.  In particular, we present results for our fiducial model (three boxes to cover a wide range of mass and spatial scales) for a range of post-processing prescriptions to model the SEDs and the effects of dust. In particular, we investigate the effects the inclusion of nebular emission, varying the IMF, and various phenomenological implementations of dust, frequently used in hydrodynamical simulations.  We considered single- and double power law scaling relations between metallicity and dust-to-metal ratio, two attenuation curves, and dust reddening by ISM alone and ISM and birth clouds. For each model, with different combinations of post processing assumptions, we fixed a single parameter $\epsilon$ for each $z$, by fitting the observed UVLF through $\chi^2$ minimization. Various predictions of these calibrated models such as (i) the B-band luminosity function, (ii) the H$\alpha$ luminosity function, (iii) UVLF of $z\sim$12.5 galaxies, (iv) the relationship between ${\rm M_{dust}}$ and ${\rm M_\star}$, (v) the relationship between \mab\ and \buv, (vi) the relationship between ${\rm M_\star}$ and the Balmer ratio, and (vii) the relationship between ${\rm E(B-V)_{star}}$ and ${\rm E(B-V)_{neb}}$ are compared with the corresponding available observations.
Our main results can be summarized as follows.
\begin{enumerate}[label=\arabic*., leftmargin=*, labelindent=0pt]
    \item {\bf Dust content and its redshift evolution:}  For a given set of parameters (both used in our simulation and in the post processing), the models by construction, are made to fit the observed UVLF by varying $\epsilon$. As it is well documented, we also need the dust reddening to play an important role in reproducing the shape of the UVLF at the high-luminosity end.  On the other hand, for a given simulation, the UVLF in  the low-luminosity end is highly sensitive to the assumed IMF.  Therefore, the redshift evolution of UVLF, at the high-luminosity end, provides vital clues on the redshift evolution of dust.    We show that a wide range of light synthesis models (including variations in IMF and nebular emission) successfully reproduce the observed UVLF of galaxies at $5\leq z \leq 10$ with an appropriate choice of $\epsilon$ (See Figures~\ref{fig:ninja_LF} and \ref{fig:ninja_lf_comp}).

        \vskip 0.05in
    Dust reddening effects are important for galaxies at $5\le z\le 8$ and not well constrained for galaxies at $z\ge 9$. The best fit $\epsilon$ values, presented in Table~\ref{tab:lf_fit_results}, indicate a decreasing trend in dust-to-metal ratio at a given metallicity with increasing redshift. However, the best fit value of $\epsilon$ we derive depends on how we implement the dust effects and can vary by up to a factor of 7 at a given $z$. For example, our fiducial model using Calzetti attenuation curve requires a dust-to-metal ratio at $z=5-6$ to be roughly 35\% of what is seen in the Milky Way for the gas of a given metallicity. This becomes $\sim$25\% for $z\sim8$ and $\le10$\% for $z\ge9$. Instead, if we use the SMC extinction curve, the derived values are $\sim$21\%, $\sim$6\% and $\le$ 5\% for $z$ = 5, 8 and $\ge 9$ respectively. Similarly, inclusion of birth cloud attenuation substantially ($\sim$ 3.5 times for $5\le z\le 8$) reduces the dust-to-metal ratio in the ISM gas compared to our fiducial model.  Therefore, additional observational constraints are needed to draw firm conclusions on the dust parameters at a given redshift.

    \vskip 0.05in
    Our birth cloud attenuation model is similar to "model-B" discussed in \citet{Vogelsberger2020}. They get $\tau_\text{dust}$ = 0.08 and 0.03 for $z$ = 5 and 8 respectively. The corresponding values in our case are $0.106\pm0.006$ and  0.065$\pm$0.007, suggesting a consistent trend.
    The differences seen in the absolute values of $\tau_\text{dust}$ could arise from differences in the simulations and/or from the UVLF data used to constrain $\epsilon$. Future millimetre observations with ALMA  will provide stronger constraints on the dust content and its physical nature in high-redshift galaxies. We plan to extend our spectral synthesis modelling to this wavelength regime in future studies.

   \vskip 0.05in 
    \item {\bf Nebular emission:} We have used the photoionization code \software{CLOUDY} to compute nebular emission associated with each star particle.  For our simulations, inclusion of nebular continuum systematically boosts the UV luminosity of high-$z$ galaxies and improves agreement with the observed UVLFs (in particular at low-luminosities).   While nebular emission has a moderate impact on the UVLF, its contribution can be appreciable for the observed B-band (or rest frame optical) luminosity function. We notice that simultaneous fitting of UVLF and B-band luminosity function at high luminosity end can be very useful in constraining the slope of the extinction curve (see section~\ref{sec:BLF}). At present, large measurement errors in the available observations prevent us from drawing such conclusions. At the low luminosity end, our predictions are closer to the observed B-band luminosity functions when we include the contribution of nebular continuum emission.

    \vskip 0.05in
    Our predicted H$\alpha$ luminosity functions,  for galaxies at $4.9\le z\le 6.7$, are higher than the observed ones (section~\ref{sec:lf_halpha}). We argue that this can be rectified by introducing (i) dust attenuation inside the ionised nebula modeled using \software{CLOUDY} (as at present dust free gas is assumed in the H~{\sc ii} regions) and/or (ii) by allowing escape of ionizing radiation from the H~{\sc ii} regions around star particles.  Both these assumptions will have direct implications on (i) the diffuse continuum emission we include in the SEDs, (ii) various derived line luminosities, and (iii) H~{\sc i} reionization and hence the damping wing produced by the \lya\ absorption. 

    \vskip 0.05in
    \item {\bf Attenuation Curves:} In this study we consider "Calzetti" attenuation and SMC extinction curves. While the calibrated UVLF for both cases fit the observational data equally well they differ in their predictions of (i) B-band luminosity function, (ii) H$\alpha$ luminosity function and (iii) \mab$-$\buv \ relationship.  The models using SMC extinction curve produce higher B-band and H$\alpha$ luminosity function (at their respective high-luminosity end) compared to the models that use Calzetti attenuation curve. Thus well measured rest-frame optical luminosity functions (in particular at the high luminosity end) can be very useful in constraining the average attenuation curve of the high redshift population of galaxies.

    \vskip 0.05in
    We also show the slope of the \mab \ vs \buv \ relationship is more sensitive to the assumed attenuation curve compared to the variations considered in the other dust parameters. In particular, the steepness of this relationship at $z=5-7$ is consistent with the average attenuation curve being closer to the SMC like extinction curve (section~\ref{sec:colormag}). However, due to the large values of $\epsilon$, data for $z>8$ galaxies do not constrain the slope of the extinction curve very well. 
    Interestingly, models using SMC extinction curve tend to produce less ${\rm M_{dust}}$ for a given ${\rm M_\star}$ compared to the same model that uses Calzetti attenuation curve (see Figure~\ref{fig:stellar-to-dust-mass-relation}). 

    \vskip 0.05in
    Independent constraints on the nature of extinction curves at different $z$ are obtained using SED fitting techniques.
    Data analysed by \citet{Markov2025} suggest that the average powerlaw slope and 2175\AA\ UV-bump strength are less at $z>5$ compared to the local universe. On the other hand, \citet{Fisher2025} have found a wide range of attenuation curves among the galaxies studied by them, the slope being systematically lower than what has been seen for SMC and slightly steeper than that of \citet{Calzetti2000}. They also found $\sim$30\% of their sample to show 2175\AA\ UV bump albeit with lower strength. \citet{Shivaei2025}, with the larger sample of galaxies, have found the trend of steeper slope at lower $A_V$  even at high-$z$ albeit the slope being systematically flatter for a given $A_V$. They suggested that the attenuation curves at $7\le z\le 9$ are even shallower than that of \citet{Calzetti2000}.

    \vskip 0.05in
    Similarly, observed IRX vs \buv\ relationship  (over a wide redshift range using ALMA and JWST observations) can provide independent constraints on the extinction curves at different redshifts. Latest measurements seem to favor Calzetti like attenuation curves for $5<z<7$ galaxies \citep[see for example,][]{Ciesla2025}. This is contrary to our conclusion based on \mab\ vs \buv\ relationship.
    Therefore, ideally one should  build a dust model that can simultaneously explain the  \mab\ vs \buv \ and IRX vs \buv\ relationships. For example, in order to produce steep relationship  \mab\ vs \buv\ relationship, while using the Calzetti extinction curve, we may need larger $A_V$ which may constrain the parameter used in our simulations.
    However, to draw firm conclusions, we have to ensure that the available observations are representative of the general population of galaxies at a given $z$.
    Alternatively, one has to impose appropriate selection biases for different observables in the simulated data while comparing with observations.
    
\vskip 0.05in
   We note that, in this work, the attenuation law is assumed to be fixed and is scaled only by the galaxy visual extinction. However, observational studies suggest that the shape of the attenuation curve itself varies with the metal enrichment of galaxies \citep{Salim2020}. Investigating such metallicity-dependent attenuation laws in spectral synthesis will be the subject of future work.
  
    \vskip 0.05in
    \item {\bf Dependence of dust-to-metal ratio with metallicity:}
    We explored different dust-to-metal scaling, including linear and broken power-law dependencies on metallicity. While linear relationship is frequently used in numerical simulations, the double power-law dependence of DTG on metallicity is also motivated by dust evolution models that consider various formation and destruction processes \citep[][]{Asano2013,  Zhukovska2014, Galliano2021} and also supported by low-$z$ observations   \citep[see][]{Remy_Ruyer_2014}. 
    The $\epsilon$ values required to fit the UVLF at a given $z$ is typically less by a factor 2-3 when we consider the broken power-law. However,
    we  notice that only observables that are useful in distinguishing between the two alternatives are the relationship between ${\rm M_{dust} - {M_\star}}$ (see Figure~\ref{fig:stellar-to-dust-mass-relation}) and dust mass functions (not shown in this work). The broken power-law scaling produces a steeper relationship and the existing upper limits on ${\rm M_{dust}}$ in the lower ${\rm M_\star}$ bins seem to favor this. More ALMA measurements, in particular for the general population of galaxies, covering a wide ${\rm M_*}$ range is needed to arrive at a firm conclusion.
    For all other observables discussed in this work the predictions of models using these broken power-law scaling relationship are nearly identical.

     \vskip 0.05in
    \item {\bf Single vs two component dust models:} In this work we consider models with dust only in the ISM (fiducial) and dust from ISM together with the unresolved dust from the birth clouds (birth cloud attenuation). In the second case considerable attenuation occurs at the sub-grid level and the value of $\epsilon$ for the general ISM is at least a factor 3 more (i.e less dust)  compared to our fiducial model. The main observables that are useful in distinguishing between the two alternatives are H$\alpha$ luminosity function, relationships between the Balmer ratio (H$\alpha$/H$\beta$) and ${\rm M_\star}$ and ${\rm E(B-V)_{neb}}$ vs ${\rm E(B-V)_{star}}$.  In the case of H$\alpha$ luminosity function the differences are mainly seen at the highest luminosity end (see Figure~\ref{fig:ninja_HaLF}). However, without additional constraints, this effect could be degenerate with the slope of the extinction curve. 
    We also notice that the effect of two component dust model is clearly visible only at the very high ${\rm M_\star}$  (i.e $>10^{10} {\rm M_\odot}$) end in the plot between Balmer ratio and ${\rm M_\star}$ (see Figure~\ref{fig:balmer_ratio}).
    As discussed in section~\ref{sec:Bal_ratio}, measurements of ${\rm E(B-V)_{neb}}$ are sparse and individual measurements do not follow any clear trend with ${\rm E(B-V)_{star}}$. However, binned measurements or measurements based on stacking do show a trend with $f\sim0.5$. Our models, considering ${\rm A_V^{BC}=2\times A_V^{ISM}}$, do not produce $f\sim0.5$ for ${\rm E(B-V)_{star}}<0.10$, however produces consistent values of $f$ at large ${\rm E(B-V)_{star}}$. This could imply that (i) the ${\rm \tau_V^{BC}}$ need not scale with ${\rm \tau_V^{ISM}}$ as assumed here and (ii) the extinction curve for the parent molecular cloud may be very different from that for the general ISM. Impacts of these assumptions will be investigated in future work. Unfortunately, in the two component dust implementation, we do not have a consistent way of computing the dust mass as we assume ${A_V^\text{BC}}$ in an ad hoc manner. Thus, we will not be able to use additional constraints form ALMA measurements to refine this model.

    \vskip 0.05in
    \item {\bf Initial Mass Function (IMF):} In hydrodynamical galaxy formation simulations  IMF plays are important role in the stellar feedback (i.e generation of winds and metal enrichment). NINJA simulations use Chabrier IMF for these calculations. For our fiducial models we use the same IMF albeit with different upper mass cut-off (i.e 300 M$_\odot$ in our case). Though not self-consistent with our simulations, we do consider top heavy IMF of Kroupa (KP300-TH) to explore how the UVLF depends on the IMF.  The "KP300-TH" IMF produces more UV photos per unit star formation rate compared to our fiducial model. Therefore, typically requires lower values $\epsilon$ (or more dust) compared to our fiducial model. While such a IMF is helpful in fitting the low luminosity end, they require more dust to be formed (nearly close to what we see in the local universe at $z\sim$5) at early universe.  We show that this issue can be mitigated 
    by allowing "KP300-TH" kind of IMF only below a metallicity threshold and Charier IMF for high metallicity. Note, forming more stars in low-mass halos will eliminate the need for having IMF that enhances the production of very massive stars. Therefore, IMF effects will be degenerate with the values of simulation parameters that control the formation of stars. Joint constraints from spatial clustering analyses and studies of galaxy morphology can alleviate this degeneracy, which we plan to address in future work.

    \vskip 0.05in
    \item {\bf The $z>10$ universe:} 
     At $z > 10$, our simulations are limited by both resolution and volume. Simulation boxes L150N2040 and L250N2050 have only one galaxy at $z\sim15$. Good number of star forming halos are detected in our L50N2040 box at $z\sim$12.5. We compared our models with the available observations.  The observed UVLF has a large scatter at a given UV luminosity. However, our fiducial model predictions, even before applying dust attenuation, are much below these observations. The situation is slightly better when we consider KP300-TH IMF and no dust attenuation.  We also note that there is no clear convergence in the UVLF between different boxes in the overlapping luminosity range. 
    Our results therefore indicate that higher-resolution simulations are required to robustly probe early galaxy populations.
\end{enumerate}

A key implication of our results is that dust plays a central role even at very high redshifts (i.e $z\sim 9$). The need for non-negligible dust attenuation to match the bright end of the UVLF suggests that dust formation and evolution must be efficient in early galaxies. However, in our approach, the exact scaling of dust with metallicity remains uncertain, and future observations (e.g., from ALMA and JWST spectroscopy) will be crucial in refining these models. We also find that nebular emission and IMF assumptions significantly influence inferred galaxy properties, particularly at the faint end. This has important consequences for interpreting star formation rates, stellar masses, and ionizing photon budgets during the epoch of reionization. The slight underproduction at the faint end of the UVLF (especially for $z\ge7$) suggests the need for an overall enhancement in the star formation rates of low-mass galaxies, which are strongly regulated by stellar wind feedback.
To explore this, we have been generating a suite of NINJA simulations by varying the parameters controlling the wind feedback (i.e $\kappa_w$ and ${\rm \sigma_{DM}^{1D}}$). Detailed analysis, such as the one presented here,  is being carried out for these simulations to understand various degeneracies between simulation parameters and parameters related to dust and IMF.

In this work, we identify a halo using the FOF and obtain the luminosities by integrating light from stars within the identified halos. Observationally, one uses images to identify galaxies. Thus, the number of galaxies identified and their luminosities will be slightly different from what we obtained now. It will be important to test how the derived $\epsilon$ changes if we use the galaxy count based on the images generated from the simulation boxes. 
Such an image based approach will  allow us to compare our predictions of galaxy morphological parameters with the observations. In particular, it will be  interesting to explore how the different feedback parameters affect the galaxy morphology at different redshifts. We have also varied our fiducial simulation with different black hole seeding prescriptions. The effect of this variation on the derived properties of galaxies will also be presented in our forthcoming papers.

\section*{Acknowledgements}
RB thanks Prof. Aseem Paranjape and Prof. Tirthankar Roy Choudhury for insightful discussions during the course of this project. RS thanks Prof. K. Subramanian for very useful discussions. NK thanks Simeon Bird for useful discussions and clarifications regarding MP-Gadget, as well as for identifying and fixing bugs encountered during the simulation campaign. NK acknowledges support from the IUCAA Associateship Programme. We acknowledge the use of the IUCAA HPC cluster Pegasus and the NISER HPC facilities for the computational work presented in this study.

\noindent
\textit{Software}: \texttt{MP-GADGET} \citep{Feng2018}, \software{CLOUDY} \citep{Cloudy_C23}\\
\textit{Python Packages} :\software{NumPy} \citep{2011CSE....13b..22V}, \software{SciPy} \citep{2020NatMe..17..261V},
\software{astropy} \citep{2013A&A...558A..33A},
\software{Matplotlib} \citep{2007CSE.....9...90H},
\software{jupyter} \citep{2016ppap.book...87K},
\software{mpi4py} \citep{mpi4py2021}.

\section*{Data Availability}
The data supporting the findings of this study, including the simulation datasets generated and analyzed herein, are available from the corresponding author upon reasonable request.

\bibliographystyle{mnras}
\bibliography{reference}

\appendix
\newpage
\section{Resolution Corrections}
\label{sec:res_cor}

In this work, we used three simulation boxes with different box lengths (i.e 50, 150 and 250 $h^{-1}$ cMpc).
Here, we quantify the variations in the physical quantities that arise from the resolution differences and explain how we correct them to get a combined result.
For the dark matter halo mass function ($\Phi({\rm M_{DM}})$), the halos, identified using FOF, are 
grouped  in linearly spaced   $\log_{10}({\rm M_{DM})}$ 
bins over the range $8 \le \log_{10}{\rm (M_{DM}/M_\odot)} \le 12$. The halo mass function   
in each of these mass bins is computed for each simulation box.
Due to resolution limits and finite volume effects, $\Phi({\rm M_{DM}})$ exhibits a sharp downturn at the low-mass end and fluctuates significantly at the high mass end due to small-number statistics.
To identify statistically reliable and complete bins for each box, we apply the following selection criteria:
(1) $\Phi({\rm M_{DM}}) > 0$ to remove empty bins;
(2) bin masses greater than the mass at which $\Phi({\rm M_{DM}})$ is maximal, to exclude the incomplete low-mass tail;
(3) halo counts per mass bin should be $> 10$, to suppress noisy fluctuations at the high-mass end.

\begin{figure}
    \centering
    \includegraphics[bb=15 5 395 260, clip=true, width=\columnwidth]{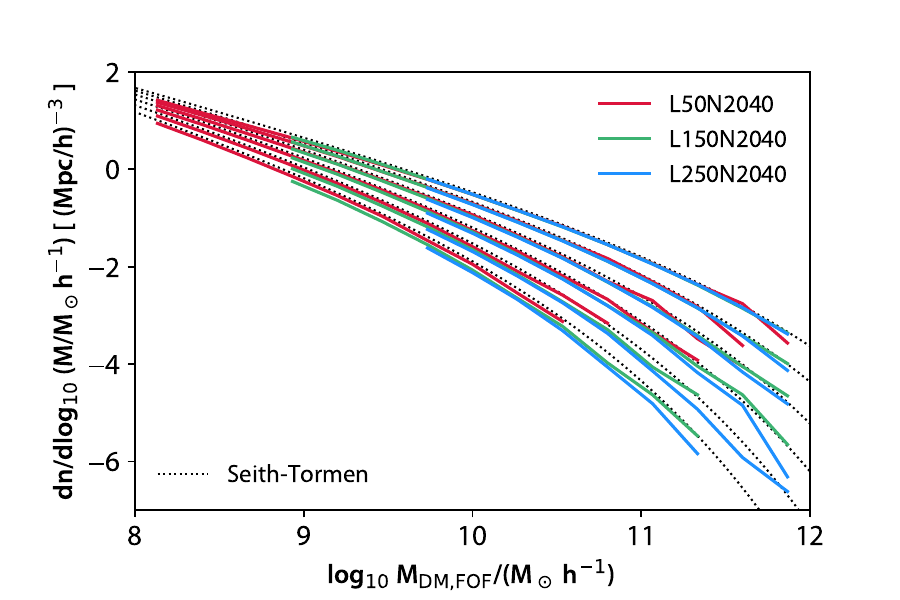}
    \caption{Redshift evolution of dark-matter Halo Mass Function, for halos identified using FoF algorithm. Results from different boxes are shown with diferent colors. The dotted black line is the standard Sheth–Tormen analytic mass function \citep{ShethTormen1999}.
    The results are shown for $z=5$  to $z=10$ from top to bottom.
    }
    \label{fig:fof_dmmf}
\end{figure}

To further remove residual incompleteness in the low mass bins, we compare the $\Phi({\rm M_{DM}})$ computed for a given box with that of the higher-resolution reference box and retain only the bins where the deviation in $\Phi({\rm M_{DM}})$ is less than $10\%$. This procedure yields a contiguous, complete, and statistically robust segment of the halo mass function for each box, as shown in Figure~\ref{fig:fof_dmmf}. For each such segment, we identify the overlapping bins with the corresponding segment of the high-resolution box. These typically consist of multiple overlapping bins with a small systematic offset. This offset is corrected by shifting halo masses, where the average offset is computed from individual bin offsets using either the mean or the median. The resulting correction is then applied uniformly to all bins.
Using this method, we find that correcting low-resolution boxes to match high-resolution ones requires a positive mass correction of $\sim 15\%$ at $z=10$, which systematically decreases to less than $1\%$ by $z=5$. The inferred correction is insensitive to whether the mean or median is used to average offsets across bins. In Figure~\ref{fig:fof_dmmf}, for comparison, we also show the Sheth \& Tormen mass function for the parameters assumed in our simulations. Since differences are minor for $z\le8$ we do not apply any completeness correction to the halo masses.
We used the halo mass as a baseline to obtain the incompleteness corrections for other quantities such as stellar mass (${\rm M_\star}$), gas mass (${\rm M_g}$), and  metal mass. In the following, we explain the procedure adopted in the case of ${\rm M_\star}$.

\begin{figure*}
    \centering
    \includegraphics[trim={0.8cm 0cm 0.8cm 0cm},clip,width=\textwidth]{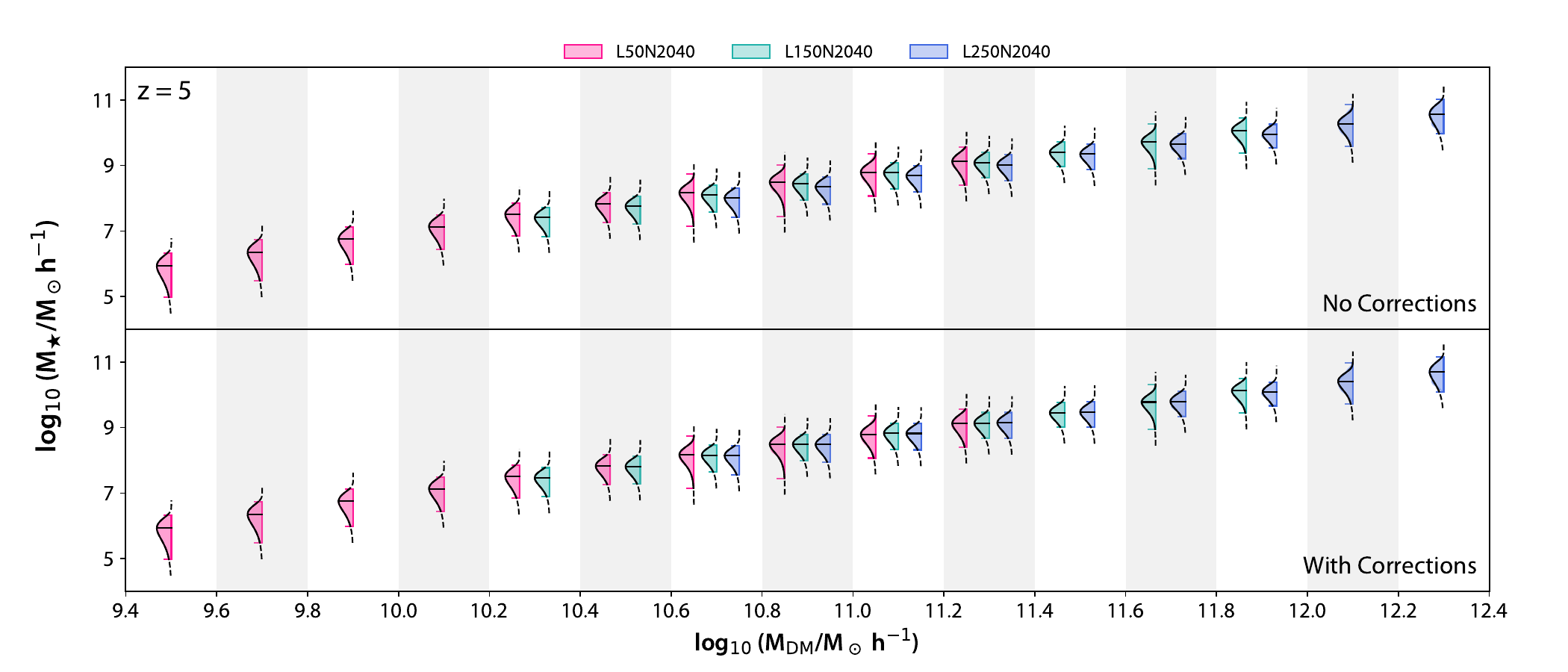}
    \caption{Correction for resolution effects in the dark matter halo mass--stellar mass relation at $z=5$. For each dark matter halo mass bin, the stellar mass distribution is shown for the three simulation boxes: L50N2040 (red), L150N2040 (green), and L250N2040 (blue). The solid black curve represents the reflected log-normal fit to the stellar mass distribution (see text for details), while the dashed black curve shows the extrapolation of the fitted form beyond the stellar mass limits in each bin. The upper panel shows the relation without any resolution correction, whereas the lower panel shows the corrected result.}
    \label{fig:ninja_dm2sm_convergence}
\end{figure*}

We examine the convergence of the distribution of ${\bf M_\star}$ for a given ${\rm M_{DM}}$, across different boxes.
Here, we assume that different boxes sample the same underlying stellar mass distribution within each  ${\rm M_{DM}}$  bin. We consider all halos that contain at least one star particle.
For each halo mass bin, the distribution of ${\rm M_\star}$ is analysed using violin plots
(colored curves in Figure~\ref{fig:ninja_dm2sm_convergence}),
and incomplete low-mass bins are excluded. High-mass bins with fewer than $100$ halos are also rejected to suppress fluctuations due to small-number statistics. Selecting halos with at least one star particle, rather than a higher threshold (e.g 16 stars), enables a better reconstruction of the low-mass tail of the stellar mass distribution. 

For bins with sufficient halo counts, the upper and lower $5$ percentiles of stellar mass are removed to reduce tail-driven fluctuations. The remaining stellar mass distribution is fitted with a reflected log-normal distribution given by,
\begin{align}
    f(M_\star)= \frac{A}{(\mu-M_\star)\sigma\sqrt{2\pi}}\exp \left(-\frac{(\ln(\mu-M_\star))^2}{2\sigma^2}\right).
\end{align}
This is shown by the solid black curve in Figure~\ref{fig:ninja_dm2sm_convergence}, while the dashed black curve represents the extrapolation of the fitted form beyond the stellar mass limits.
We determine the peak of the fitted stellar mass distribution in each halo mass bin for all simulation boxes, corresponding to the mode (i.e. the most probable value), dented by ${\rm M_\star^{peak}}$. For bins where at least two boxes overlap (after excluding statistically insignificant bins as described above), we compute the offset in the mode relative to the nearest higher-resolution box. 

These offsets (in the logarithmic scale) are typically negative for lower-resolution boxes, as seen in the upper panel of Figure~\ref{fig:ninja_dm2sm_convergence}. For each lower-resolution box, we then compute the mean offset across bins, requiring a minimum of two overlapping bins with the corresponding higher-resolution box. This average offset is subsequently applied as a positive correction to the stellar masses, thereby compensating for resolution effects (see lower panel of Figure~\ref{fig:ninja_dm2sm_convergence}). Since the analysis is performed in logarithmic space, this additive correction corresponds to a multiplicative factor greater than unity in linear stellar mass for lower-resolution simulations. These corrections are applied in a nested fashion, successively matching each box to the next higher-resolution one, and ultimately to the highest-resolution simulation. At redshift $z=5$, we find multiplicative correction factors of $1.132$ and $1.357$ for the L150N2040 and L250N2040 boxes, respectively, with respect to our highest resolution simulation box L50N2040.

After correcting all boxes such that the most probable stellar mass in each halo mass bin matches that of the highest-resolution box, we compare the fitted parameters $\mu$ and $\sigma$ as functions of dark matter halo mass. The third parameter, $A$, represents the amplitude of the distribution and is non-physical, as it depends on the number of violins plotted within a bin due to the violin-plot-based extraction. For the reflected log-normal distribution, the most probable value (mode of distribution) and variance are given by
\begin{align}
    \text{Mode} &= \mu - \exp\left(- \sigma^2\right)\\
    \text{Variance} &= (\exp(\sigma^2)-1)\exp(\sigma^2)
\end{align}

We compile these relations over multiple redshifts ($5 \le z \le 8$), requiring at least two overlapping bins for each pair of boxes. The most probable values show good convergence across boxes. Curves from different boxes are stitched using weights proportional to the halo count in each bin \citep[refer section 2.1 of][]{Kannan2022MTNG}
\footnote{All individual curves obtained from our three simulation boxes are stitched together following the methodology described in this paper.}. 

\begin{figure}
    \centering
    \includegraphics[trim={0cm 0cm 0cm 1.8cm},clip,width=\linewidth]{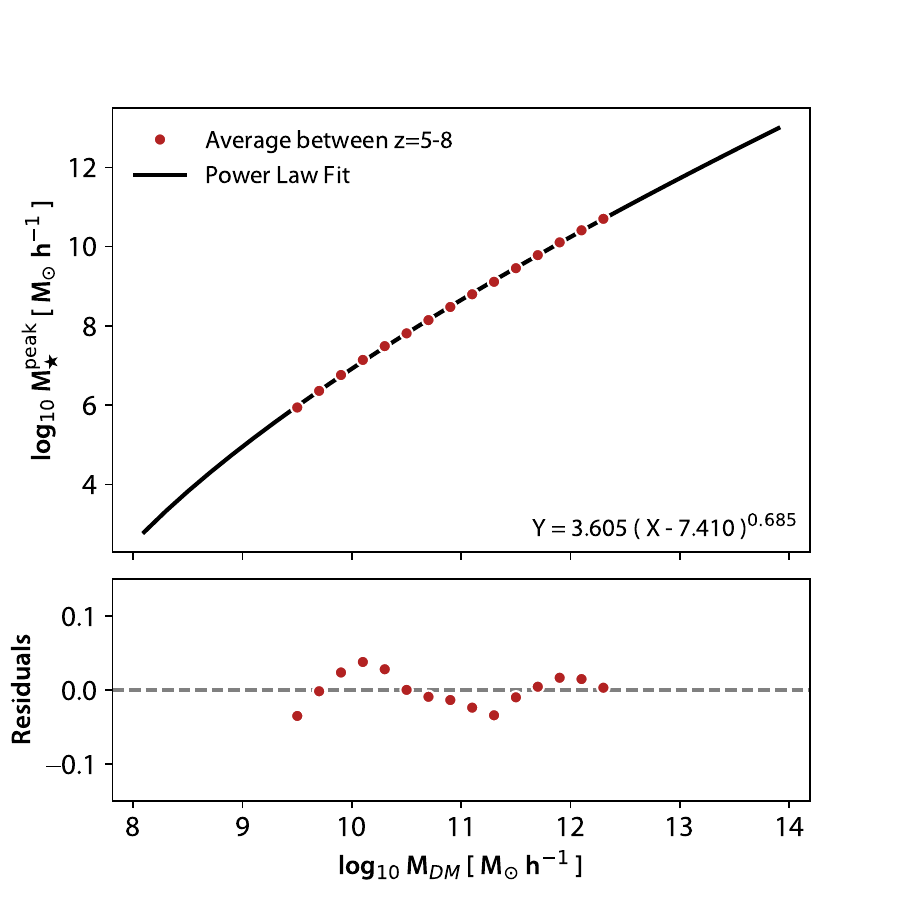}    
    \caption{{\it Top panel}: Power-law fit to the relation between dark matter halo mass and the peak of the stellar mass distribution over the redshift range $z=5-8$. To construct the fit, redshift-dependent offsets relative to $z=5$ were first removed from the binned data by applying a global correction, determined from the average offset across all halo mass bins at each redshift. This procedure removes systematic redshift evolution, yielding a redshift-independent halo mass–stellar mass relation. The corrected binned data from the four redshifts were then combined, and the mean value in each halo mass bin was computed from the corresponding four data points, shown as red dots. These averaged points were subsequently used to derive the best-fitting power-law relation. {\it Bottom panel}: The residual deviations from the fitted relation as a function of halo mass.}
    \label{fig:dm2sm_fit_relation}
\end{figure}

Relative to $z=5$, the redshift-dependent offset is systematic toward lower stellar masses but remains below $2\%$ even at $z=8$. The standard deviation increases toward lower halo masses and higher redshifts, but remains two orders of magnitude smaller than the most probable value, indicating a tight distribution. We remove the small redshift-dependent offset in the most probable stellar mass–halo mass relation by applying an average shift to align all redshifts. The relations are then averaged in each bin over $5 \le z \le 8$ to obtain a single matched dataset $Y$. This combined relation is well described by a power law (Figure~\ref{fig:dm2sm_fit_relation}), with residual deviations below $1\%$, given by
$Y = 3.605 \left(X - 7.410\right)^{0.685}$ where $X=\log_{10}{M}_\text{DM}$. All calculations were performed using log-scaled masses in units $M_\odot/\text{h}$.

\begin{figure}
    \centering
    \includegraphics[trim={0cm 0cm 0cm 0cm},clip,width=\linewidth]{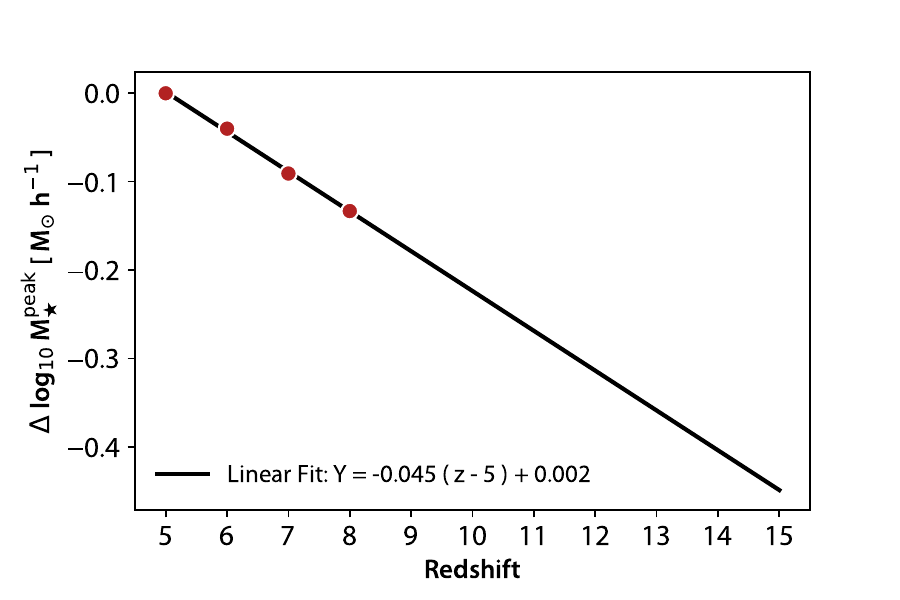}
    \caption{The red points show the redshift-dependent offsets in the relationship between the halo mass and peak stellar mass, used to remove systematic redshift evolution, as described in Figure~\ref{fig:dm2sm_fit_relation}. A linear fit is used to characterize the evolution of these offsets with redshift.}
    \label{fig:dm2sm_redshift_offset_fit}
\end{figure}

We next model the redshift evolution of the stellar mass–halo mass relation by quantifying its offset relative to the $z=5$ reference relation. For this purpose, we compute the average offset at each redshift, considering $z = 5, 6, 7,$ and $8$, requiring a minimum of two overlapping halo mass bins between simulation boxes to ensure a consistent comparison. The redshift dependence of this offset is modelled using a linear fit (Figure~\ref{fig:dm2sm_redshift_offset_fit}). This fitted relation is subsequently extrapolated to higher redshifts, up to $z=15$, allowing us to compensate for resolution limitations and increasing incompleteness in the high-redshift regime. The fit relation is given by $Y=-0.045(z-5)+0.002$. Combining both fits we have the calibrated dark matter halo mass to stellar mass relation:
\begin{align}
    \log_{10}\left(\frac{M_\star^{\rm peak}}{M_\odot/h}\right) = & \left[0.002-0.045(z-5)\right] \nonumber \\
    & \times \left[3.605\left(\log_{10}\left(\frac{M_\text{DM}}{M_\odot/h}\right)-7.410\right)^{0.685}\right]
\end{align}

At $z = 9$ and 10, where fewer than two halo mass bins overlap between simulation boxes, we employ the previously calibrated redshift-dependent relation to correct for resolution effects. For each box, we first determine the offset required to match the calibrated relation within each halo mass bin and compute the average offset across all valid bins. These average offsets are then compared between boxes, using the nearest higher-resolution simulation as the reference. The difference in average offset between two boxes defines the correction factor required to bring the lower-resolution box into agreement with the higher-resolution one. This procedure is applied in a nested manner across all simulations, ultimately matching all boxes to the highest-resolution run, L50N2040.

We also calibrate the stellar mass–metal mass relation over $5 \leq z \leq 9$ across simulation boxes, adopting the same bin-wise distribution-based methodology used for the halo mass–stellar mass relation. The resulting corrections for the lower-resolution boxes are incorporated into the light synthesis and column density calculations. Following this procedure, we observe good convergence in the relations between dark matter halo mass and both intrinsic and attenuated magnitudes. The intrinsic magnitude shows a higher degree of convergence, while the remaining small discrepancy in the attenuated magnitude is primarily driven by residual differences in column density convergence. We further note that this residual discrepancy in the attenuated magnitude can be amplified under models that assume higher average dust content, such as KP300-TH IMF model.

\section{Comparison of UVLF with that from other Simulations}
\label{sec:appendix-comp-other-sim}
In this section, we compare the calibrated UVLF from our fiducial model (discussed in Section~\ref{sec:UVLF}) with those from different cosmological hydrodynamical simulations focused on galaxy formation at high-redshifts ($z \ge 5$).   We provide brief descriptions of how these simulations differ in terms of implementation of physical processes and post processing tools.

We consider large-volume simulations from the GADGET family, including \BlueTides \citep{Feng2016, Wilkins2017},
and \ASTRID \citep{Bird2022}, as well as simulations from the AREPO family, namely \IllustrisTNG \citep{Pillepich2018, Vogelsberger2020}, \THESAN \citep{Kannan2022THESAN}, and \MilleniumTNG \citep{Kannan2022MTNG,HernandezAguayo2023,Pakmor2023}. 
In addition, we consider cosmological zoom-in simulations, including \FIRETWO \citep{Ma2018}, performed with the GIZMO code; \FLARES \citep{Vijayan2021,Lovell2021}, based on the \EAGLE\ parent simulation \citep{Schaye2015,Crain2015,Trayford2020}; and \THESANZOOM \citep{Kanna2025}, which adopts initial conditions from the \THESAN\ parent box. In  Figure~\ref{fig:compare_ninja}, we provide box side length vs particle per dimension plot for  most of these models.
We further compare to \UniverseMachine \citep{Behroozi2019}, an empirical framework that populates dark matter halos from the Bolshoi–Planck and MDPL2 simulation using observationally constrained relations. 

\begin{figure*}
    \centering
    \includegraphics[width=\textwidth]{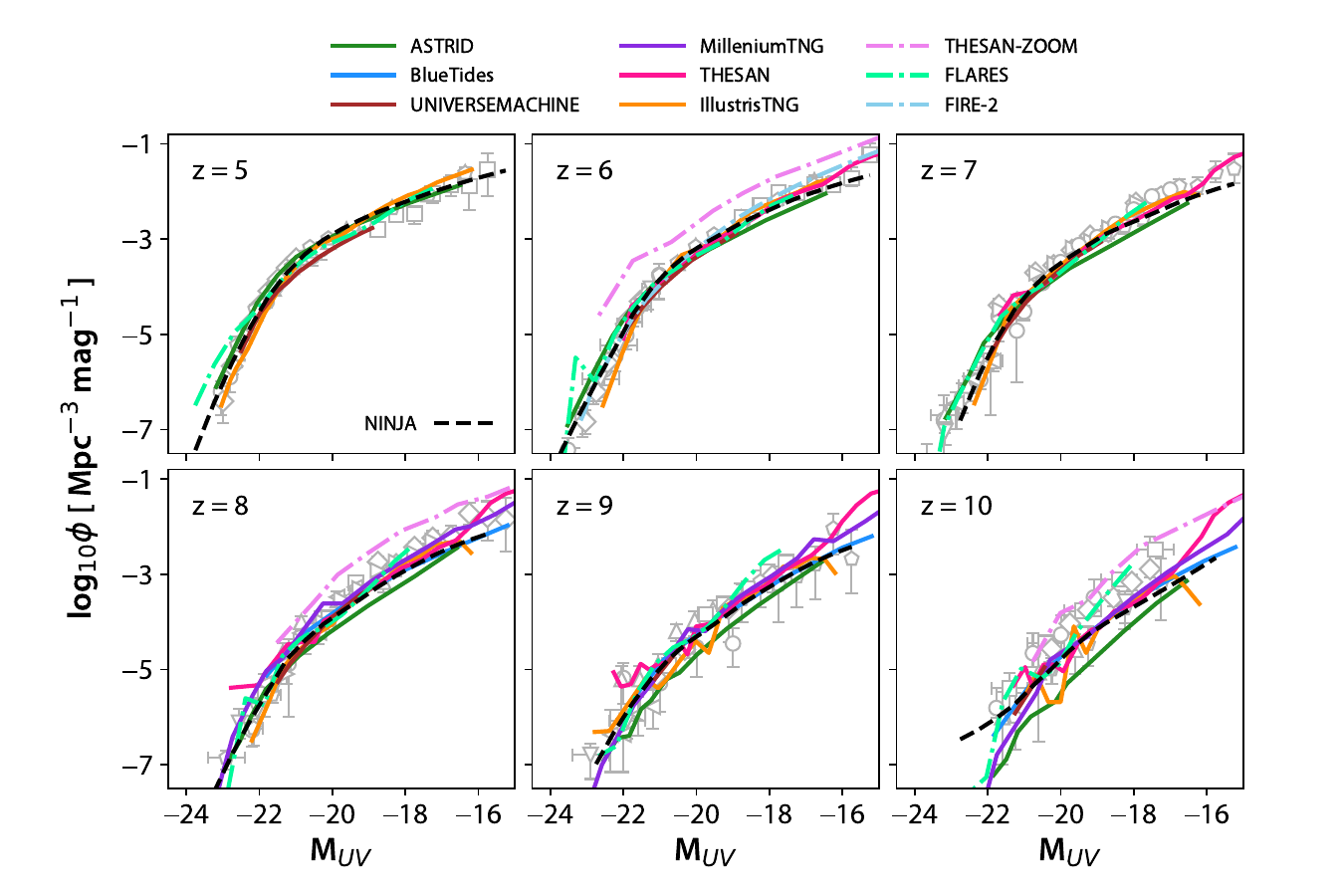}
    \caption{Comparison of the redshift evolution of the ultraviolet luminosity function (UVLF) from the NINJA simulation (shown by dashed black curves) with other cosmological simulations (solid colored curves) and zoom-in simulations (dash-dotted colored curves). All cosmological simulations show excellent agreement in the UVLF over the redshift range $5\le z \le 7$, as they are calibrated to reproduce low-redshift observational constraints. The small offsets between different models arise from variations in the underlying physical prescriptions, as discussed in the text. On average, the zoom-in simulations provide a better match to the UVLF, particularly at the faint end. However, the THESAN-ZOOM model, which is calibrated to high-redshift observations, tends to overpredict the UVLF at lower redshifts.}
    \label{fig:uvlf-compare}
\end{figure*}

The observed luminosity functions and UVLF predicted by all the simulations mentioned above, over the redshift range $5\le z\le 10$, are summarized in Figure~\ref{fig:uvlf-compare}. Recall that \NINJA\ is run using \texttt{MP-GADGET}, the same code used for \ASTRID, which itself builds upon \BlueTides\ with improved prescriptions for stellar feedback. Our best fit UVLF closely follows that of \BlueTides\ for $z\ge8$. On the other hand, \ASTRID\ simulations under predict UVLF compared to  the NINJA simulations for $z\ge7$ especially at the low-luminosity end.  The difference could come from the IMF adopted  (i.e Salpeter IMF with an upper mass cutoff of $300\,{\rm M_\odot}$), not including the contribution of nebular emission to the continuum and not allowing for the redshift evolution of dust-to-metal (or dust-to-gas) ratio with redshift in \ASTRID. On the other hand, \BlueTides\ uses the simple relationship between the star formation rate and \mab\ without performing the full spectral synthesis.  
It is also evident from the figure that our predicted UVLF closely follow the predictions of AREPO family of simulations over the full redshift range considered here. As discussed in Section~\ref{sec:simulations}, unlike our models, these simulations allow for the redshift dependent $\sigma_\text{DM}^{1D}$ and have a minimum velocity floor for wind injection velocity $v_w$. Thus, it appears that, to some extent, the differences in the feedback process can be compensated by slight differences in the assumed IMF and dust properties while reproducing the UVLF.

UVLFs predicted by the \THESANZOOM\ simulations are available for $z=6$, 8 and 10. Clearly, they differ appreciably compared to the predicted UVLF of NINJA. It is also interesting to note that while \THESANZOOM\ simulations fit the observed UVLF well at $z\sim10$, they over-produce at lower redshifts. Our predicted UVLF closely follows the UVLF predicted by the \FLARES\ simulations. However, \FLARES\  simulations produce the H$\alpha$ luminosity function much closer to observations than those of \NINJA. We believe it is mainly due to the way H~{\sc ii} regions are modeled while computing the nebular emission. \citet{Vijayan2021} include Orion type dust grains in the H~{\sc ii} regions while we do not include grains in our calculations.

We also wish to printout the importance of how we define a galaxy. The galaxy identification can be done either by using the identified halos (as done here) or by using simulated images and using techniques similar to those used by observers.
\BlueTides\  constructs mock two-dimensional images of the simulation volume (without instrumental noise) and applies SEP, a Python implementation of \texttt{SourceExtractor}, to identify galaxies. They report that SEP yields up to $\sim 20\%$ more objects at the faint end ($M_{\mathrm{UV}} > -18$) and $\sim 20\%$ fewer objects at the bright end ($M_{\mathrm{UV}} < -22$) compared to the Friends-of-Friends (FOF) halo finder at $z=8$ \citep[see Figure~6 of][]{Feng2016}. This discrepancy arises because massive FOF halos can be decomposed into multiple smaller clumps by SEP. 
Subsequently, \ASTRID\ compared the ultraviolet luminosity functions (UVLFs) constructed using \texttt{SUBFIND} and FOF, finding good agreement between the two. This is primarily because, at these redshifts, the UV luminosity is dominated by the central galaxy within each FOF group. Motivated by this, \NINJA\ adopts the FOF halo finder with a minimum threshold of 16 star particles to construct galaxy catalogs, while deferring the use of image-based identification methods to future work. In contrast, \IllustrisTNG\ employs a two-stage procedure in which FOF groups are first identified, followed by \texttt{SUBFIND} to locate bound substructures. Galaxies are then selected based on a stellar mass threshold of $m_\star = 100\,m_b$ within twice the stellar half-mass radius, where $m_b$ is the baryonic mass resolution. A similar FOF+SUBFIND pipeline is also adopted by \EAGLE, \THESAN, and \MilleniumTNG. On the other hand, \UniverseMachine\ relies on dark matter halos identified using the \textsc{ROCKSTAR} algorithm \citep{Behroozi2013}. The parent simulation used in \FIRETWO\ employs the \textsc{AHF} halo finder to select target halos for zoom-in re-simulations.

The generation of intrinsic ultraviolet emission varies substantially across simulations, reflecting differences in stellar population synthesis (SPS) models, stellar initial mass functions (IMFs), and assumptions regarding stellar evolution, metallicity interpolation, and the treatment of nebular emission. While some simulations, such as \BlueTides, adopt empirical star formation rate–based prescriptions to estimate UV luminosities, others compute stellar emission explicitly using SPS frameworks such as BPASS (\ASTRID, \THESAN, \FIRETWO, \FLARES) or FSPS (\IllustrisTNG), or estimate UV luminosities indirectly through regression-based approaches (\MilleniumTNG). Similarly, dust attenuation is modeled using a broad range of prescriptions, spanning empirical conversions between intrinsic and observed UV magnitudes to physically motivated models based on metal or gas column densities along the line of sight. Several simulations calibrate attenuation prescriptions at a fixed redshift and extrapolate them across cosmic time, while others incorporate additional birth-cloud attenuation or resolve sightline-dependent variations through ray tracing.

Despite substantial differences in simulation methodologies -- including the underlying code families, feedback implementations, halo identification techniques, prescriptions for intrinsic light generation, and assumptions regarding dust attenuation -- it is evident from Figure~\ref{fig:uvlf-compare} that most simulations are calibrated to reproduce observational constraints at lower redshifts ($z \sim 5$--$7$) with reasonable agreement. Furthermore, with the exception of \THESANZOOM\ and \ASTRID\ (see discussion above), the simulations exhibit remarkably similar evolution of the UV luminosity function (UVLF) across redshift. This convergence, despite differing physical assumptions and modeling strategies, highlights the limited constraining power of the UVLF alone in distinguishing between galaxy formation models. Consequently, robust characterization of high-redshift galaxy properties will require moving beyond UVLF-based comparisons toward additional spectral and observational diagnostics, an aspect that should be more carefully addressed in the next generation of simulations.

\section{Scaling with Halo Mass and Stellar Mass in NINJA simulation}
\label{sec:ninja-scaling-relations}

\begin{figure}
    \centering
    \includegraphics[width=\linewidth]{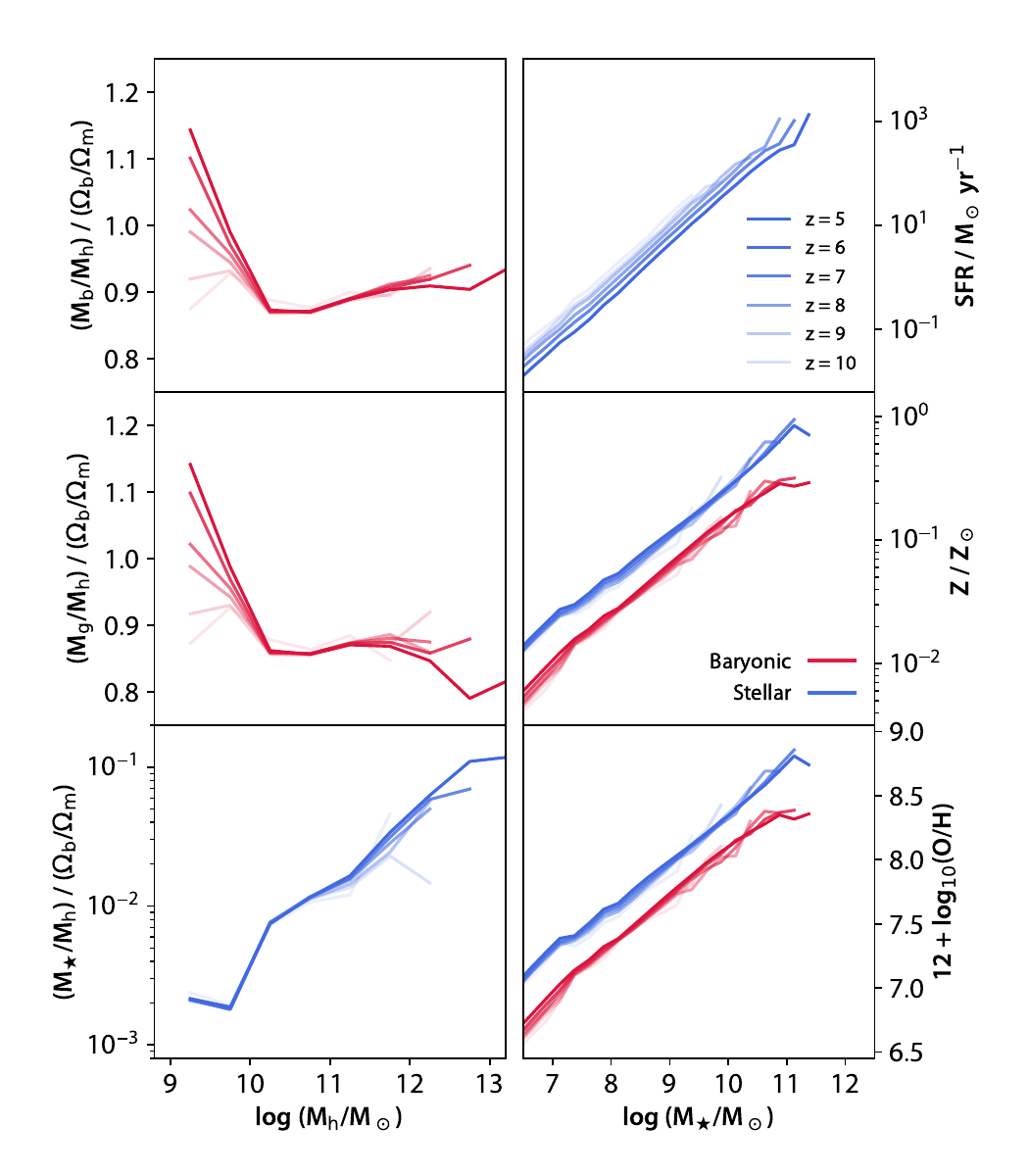}
    \caption{Median scaling relations in the NINJA simulation boxes over the redshift range $5\le z\le 10$. The left-hand panels show scaling relations with respect to Friends-of-Friends (FOF) halo mass, while the right-hand panels show relations with respect to stellar mass. The top and middle left panels show the baryon and gas mass fractions, respectively, normalized by the cosmic baryon fraction. The bottom left panel shows the stellar mass fraction, normalized in the same way, representing the star formation efficiency. The top right panel shows the instantaneous star formation rate (computed from star-forming gas particles) as a function of stellar mass, corresponding to the galactic main sequence. The middle and bottom right panels show the stellar mass--metallicity relation, expressed in terms of oxygen abundance, for the gas phase (labelled \textit{baryonic}) and stellar phase (labelled \textit{stellar}), shown in red and blue, respectively. In all panels, the curves progressively fade with increasing redshift.}
    \label{fig:ninja-scalings}
\end{figure}

In Figure~\ref{fig:ninja-scalings}, we plot scaling relations as a function of halo mass (left panels) and stellar mass (right panels), where all trends are represented by the median of the underlying distributions.
In the lower-left panel, the star formation efficiency remains approximately constant within $\sim 10\%$ across all redshifts, and increase with halo mass, rising from $\sim 0.002$ at $\log M_{\mathrm{DM}} \approx 10$ to $\sim 0.1$ at $\log M_{\mathrm{DM}} \approx 13$. At higher halo masses, the star formation efficiency is generally expected to peak and subsequently decline, as commonly found in galaxy formation studies \citep[For example, see Figure~11 of][]{Pillepich18}; however, the halo mass range probed by our simulations does not extend sufficiently into this regime. Correspondingly, the gas fraction (middle-left panel) and the total baryon fraction (gas + stars + black holes; top-left panel) remain close to $\sim 90\%$.

The right panels show scaling relations with stellar mass. The top-right panel presents the evolution of the galactic main sequence (stellar mass versus star formation rate). The middle-right panel shows the relation between gas-phase (baryonic) and stellar-phase metallicities as a function of stellar mass, where we assume a solar metallicity of $Z_\odot = 0.02$. The lower-right panel shows the corresponding relation for oxygen abundance, used here as a proxy for metallicity, for both gas and stellar phases. We find that all these quantities exhibit tight correlations with stellar mass. There is minimal evolution in both metallicity and oxygen abundance at fixed stellar mass across redshift. Furthermore, the stellar-phase metallicity and oxygen abundance are systematically higher than those of the gas phase. For the galactic main sequence, the median relation shows a slight downward shift with increasing stellar mass.

\section{Binning Observed data for \texorpdfstring{$M_{\rm AB}-\beta_{\rm UV}$}{MAB-betaUV}}
\label{sec:obs_data}

For observational datasets providing magnitude ($M_\mathrm{AB}$) and UV slope ($\beta_\mathrm{UV}$), some studies present binned measurements, whereas others report unbinned individual object data. For the unbinned samples, we first apply a redshift selection mask and divide the data into 3, 2, or 1 chunks, depending on whether the sample contains more than 40, 20, or 5 objects in total, respectively. We symmetrize the measurement uncertainties, when applicable, by averaging the positive and negative error estimates. To prevent unrealistically small uncertainties, we impose a lower limit of 0.03 on the error in $\beta$ for any individual source. For each chunk, we compute the weighted mean value of $\beta$ using inverse variance weighting:
\begin{align}
       \bar{\beta} = \frac{\sum_{i=1}^{N} \beta_i/\sigma_i^2}{\sum_{i=1}^N 1/\sigma_i^2}
\end{align}
The corresponding uncertainty in the weighted mean is given by
\begin{align}
        \Delta \bar{\beta}=\left(\frac{\sum_{i=1}^{N}\left[{(\beta_i-\bar{\beta})^2}/{\sigma_i^2}\right]}{(N-1)\sum_{i=1}^N 1/\sigma_i^2}\right)^{0.5}
\end{align}
We further compute the average magnitude within each chunk to reduce the dataset to a single representative binned point expressed as $(\bar{M}_\mathrm{AB}, \bar{\beta} \pm \Delta \bar{\beta})$. {These  binned measurements are shown in Figure~\ref{fig:ninja_colmag}.}

\label{lastpage}
\end{document}